\documentclass[prd,showpacs,superscriptaddress,twocolumn,
floatfix,preprintnumbers,altaffilletter]{revtex4}
\usepackage{graphicx,url,amssymb,amsmath,longtable,rotating,color,units, 
wasysym,subfigure,epsfig,multirow}
\usepackage{epstopdf}
\usepackage{pdflscape}
\usepackage{makecell}

\newcommand{\beq}{\begin{equation}}
\newcommand{\eeq}{\end{equation}}
\newcommand{\bdm}{\begin{displaymath}}
\newcommand{\edm}{\end{displaymath}}

\begin{document}

\title{Implications of dedicated seismometer measurements on Newtonian-noise cancellation for Advanced LIGO}
\author{M. W. Coughlin}
\affiliation{Division of Physics, Math, and Astronomy, California Institute of Technology, Pasadena, CA 91125, USA}
\author{J. Harms}
\affiliation{Gran Sasso Science Institute (GSSI), I-67100 L’Aquila, Italy}
\affiliation{INFN, Laboratori Nazionali del Gran Sasso, I-67100 Assergi, Italy}
\author{J. Driggers}
\affiliation{LIGO Hanford Observatory, Richland, WA, 99352, USA}
\author{D. J. McManus}
\affiliation{OzGrav, Australian National University, Research School of Physics and Engineering, Canberra, Australian Capital Territory 2601, Australia}
\author{N. Mukund}
\affiliation{Inter-University Centre for Astronomy and Astrophysics (IUCAA), Post Bag 4, Ganeshkhind, Pune 411 007, India}
\author{M. P. Ross}
\affiliation{Department of Physics, University of Washington, Seattle, WA 98195, USA}
\author{B. J. J. Slagmolen}
\affiliation{OzGrav, Australian National University, Research School of Physics and Engineering, Canberra, Australian Capital Territory 2601, Australia}
\author{K. Venkateswara}
\affiliation{Department of Physics, University of Washington, Seattle, WA 98195, USA}

\begin{abstract}
Newtonian gravitational noise from seismic fields will become a limiting noise source at low frequency for second-generation, gravitational-wave detectors. It is planned to use seismic sensors surrounding the detectors' test masses to coherently subtract Newtonian noise using Wiener filters derived from the correlations between the sensors and detector data. In this work, we use data from a seismometer array deployed at the corner station of the LIGO Hanford detector combined with a tiltmeter for a detailed characterization of the seismic field and to predict achievable Newtonian-noise subtraction levels. As was shown previously, cancellation of the tiltmeter signal using seismometer data serves as the best available proxy of Newtonian-noise cancellation. According to our results, a relatively small number of seismometers is likely sufficient to perform the noise cancellation due to an almost ideal two-point spatial correlation of seismic surface displacement at the corner station, or alternatively, a tiltmeter deployed under each of the two test masses of the corner station at Hanford will be able to efficiently cancel Newtonian noise. Furthermore, we show that the ground tilt to differential arm-length coupling observed during LIGO's second science run is consistent with gravitational coupling.
\end{abstract}


\maketitle

Detections of gravitational waves (GWs) from compact binaries such as binary black holes \cite{AbEA2016a,AbEA2016e,AbEA2017a,AbEA2017c} and binary neutron stars \cite{AbEA2017b} can be facilitated by improving low-frequency sensitivity of GW detectors. In particular, detection of higher mass mergers (and at higher rates) is possible as the low-frequency sensitivity improves \cite{HaMa2018}. In addition, increasing sensitivity at low frequencies can significantly improve our ability to estimate certain signal parameters such as the individual masses of the two compact objects and lead to more stringent tests of general relativity \cite{SaEA2012,LyEA2015}. 

One of the major noise contributions below 30\,Hz comes from terrestrial gravity fluctuations, also known as Newtonian noise (NN) \cite{Sau1984,Har2015}. These gravity fluctuations are predominantly from two sources:  density perturbations in the atmosphere, or from seismic fields. Seismic surface fields are predicted to dominate the NN contribution \cite{DHA2012}, although recent measurements at Virgo show that the atmosphere can be important as well \cite{FiEA2018}. While the average NN is likely to lie below the instrumental noise of the Advanced LIGO and Virgo detectors, at times of higher environmental noise, it can dominate \cite{DHA2012}.

It was proposed to mitigate NN by monitoring the environmental fields with sensor arrays \cite{Cel2000}. Since site-characterization measurements suggest that seismic fields at the LIGO sites are dominated by surface Rayleigh waves in the LIGO NN band between 10\,Hz and 20\,Hz from local sources \cite{CoMu2016}, NN mitigation can be achieved by deploying a surface array around each test mass monitoring vertical ground displacements \cite{DHA2012}.

\begin{figure*}[ht!]
\includegraphics[width=3.8in]{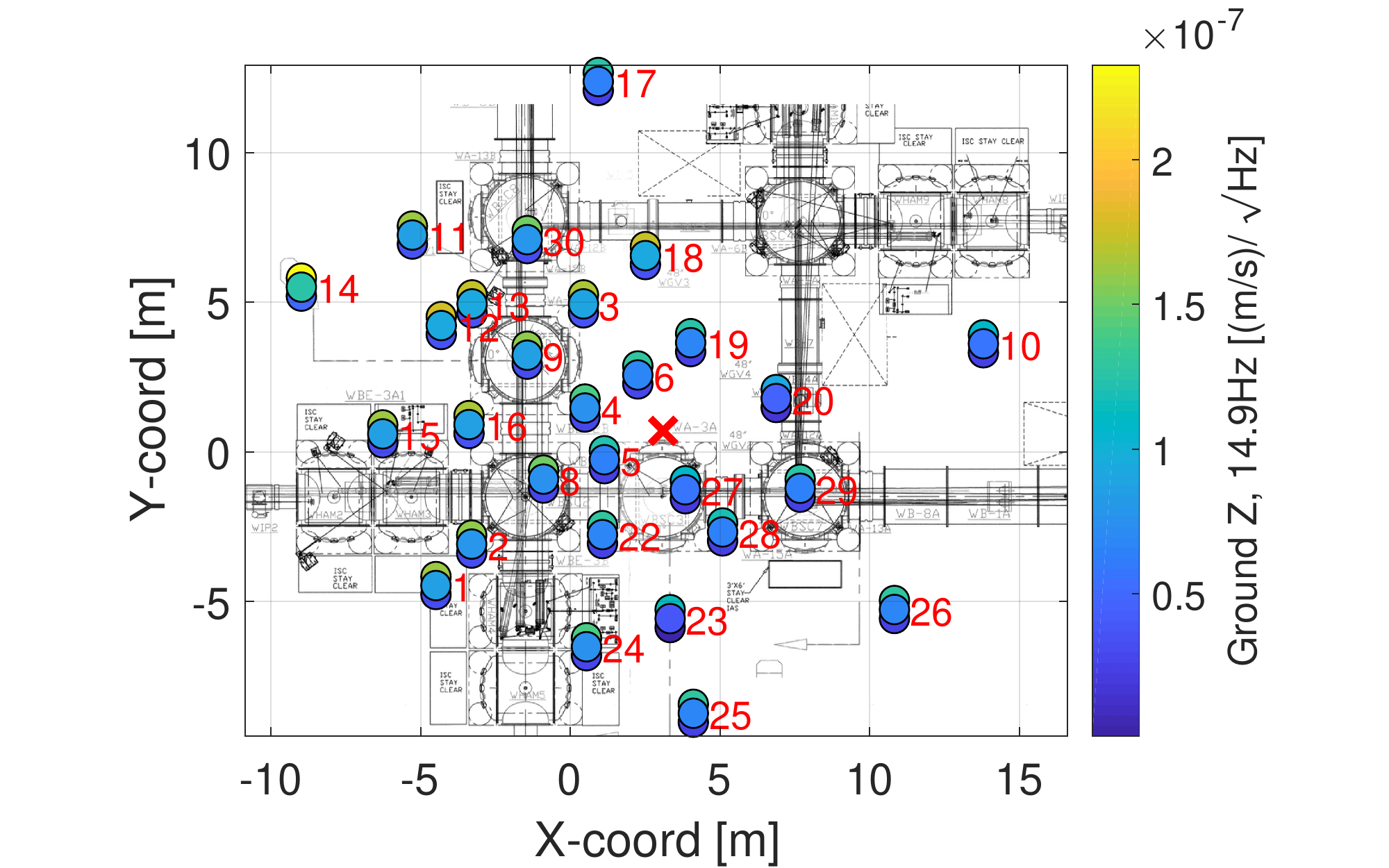}
\includegraphics[width=3.2in]{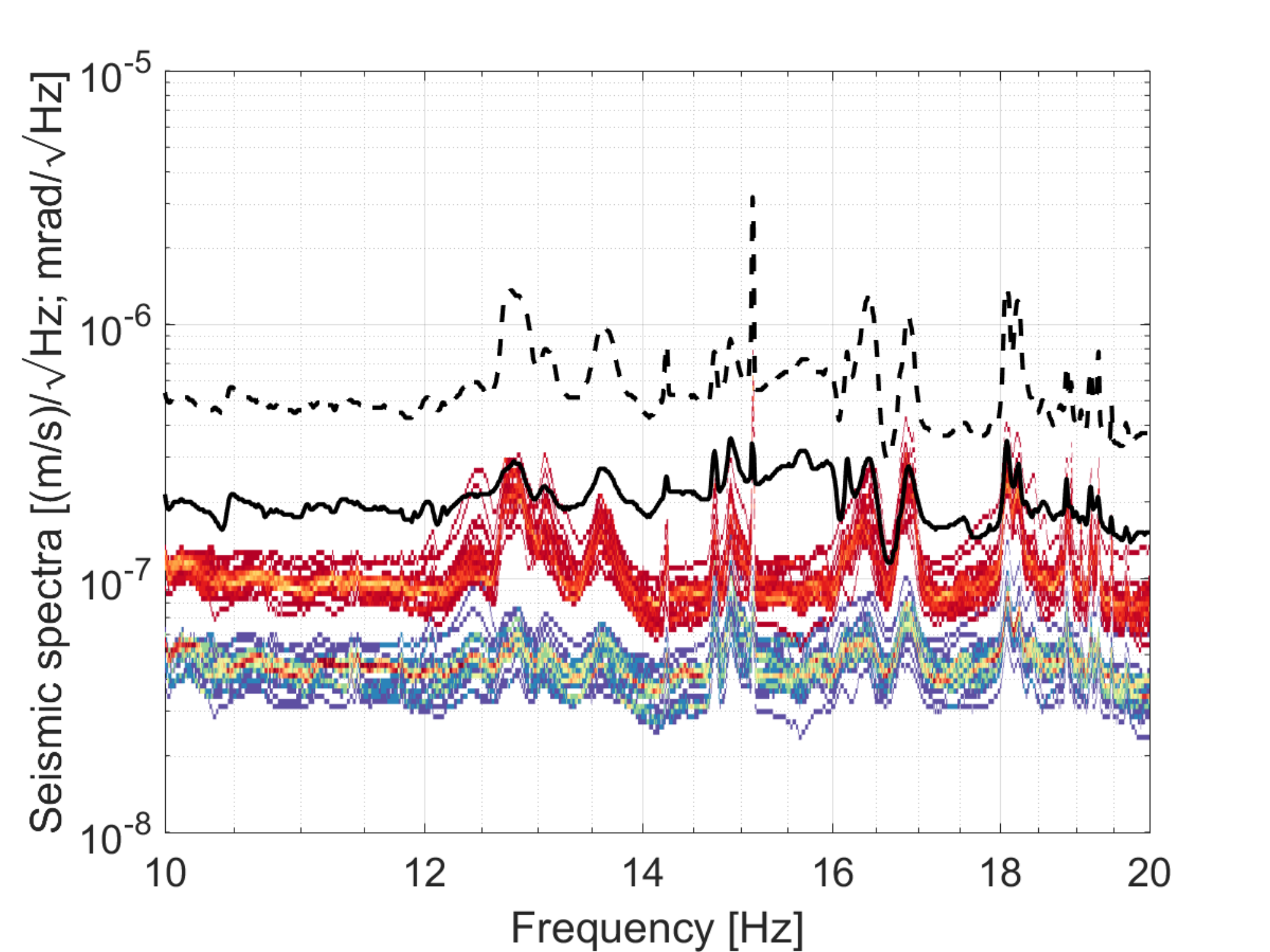}
\caption{On the left is the layout of the instrument floor at the corner station of the LIGO Hanford Observatory, with the location of the L-4C's indicated by the coloured circles and its number. For each seismometer, the 10th, 50th, and 90th percentiles of the square-root of power spectral densities (PSDs) at 15\,Hz are indicated from bottom to top with the colored circles. The red cross marks the location of the tiltmeter during the first months of the O2 science run. On the right are histograms of the 50th and 90th percentiles of seismic spectra collected from all seismometers in units $\rm(m/s)/Hz^{-1/2}$. The black lines are the 50th (solid) and 90th (dashed) percentiles of the tiltmeter spectra in units $\rm mrad/Hz^{-1/2}$.}
\label{fig:array}
\end{figure*}

The conventional approach of NN subtraction is to create Wiener filters using data from sensor arrays, similar to feed-forward cancellation schemes already used with other detector noise \cite{GiEA2003,DrEA2012,DeEA2012}. Previous work was concerned with optimizing the placement of seismometers using high-dimensional samplers minimizing the expected noise residuals \cite{DHA2012, CoMu2016}. Wiener filters are typically constructed using observed correlations between sensors, although models of the seismic field can be employed as well \cite{CoMu2016}. Harms and Venkateswara \cite{HaVe2016} also showed that a single seismic tiltmeter can be used to strongly reduce NN only limited by tiltmeter self-noise, provided that the seismic field is accurately represented by plane-wave models. 

\begin{figure}[t]
 \includegraphics[width=3.5in]{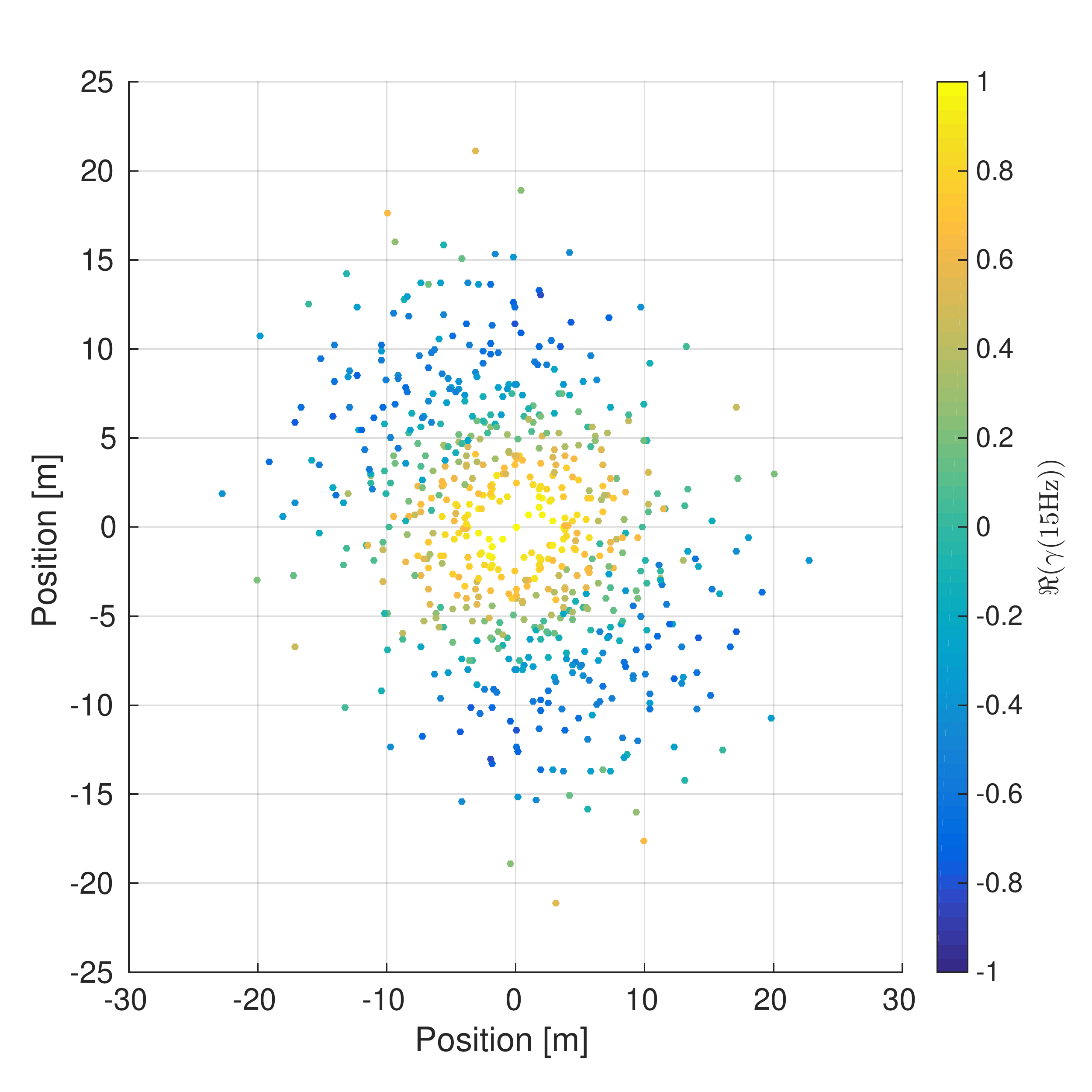} 
 \caption{
 The measured correlation function for the LIGO Hanford corner station at 15\,Hz.}
 \label{fig:coh}
\end{figure}

In this Letter, we present first results of the cancellation of a tilt signal in the frequency range 10\,Hz --- 20\,Hz measured by a compact beam-rotation sensor \cite{HaVe2016}, and give a detailed characterization of the seismic field for the purpose of NN cancellation. We first show that the seismic field is approximately homogeneous and dominated by Rayleigh waves, which is important as seismic fields with a mixture of wave types would be very difficult for any NN subtraction \cite{Har2015}. We then implement an optimal subtraction scheme using the array of seismometers as input to a Wiener filter and the tiltmeter as target channel. The investigation with the tiltmeter as a target channel is important, because ground tilt along the direction of the detector arm is fully coherent with NN from plane Rayleigh waves, and therefore is the best available proxy for testing NN cancellation schemes \cite{Har2015}. 

\begin{figure}[t]
 \includegraphics[width=3.5in]{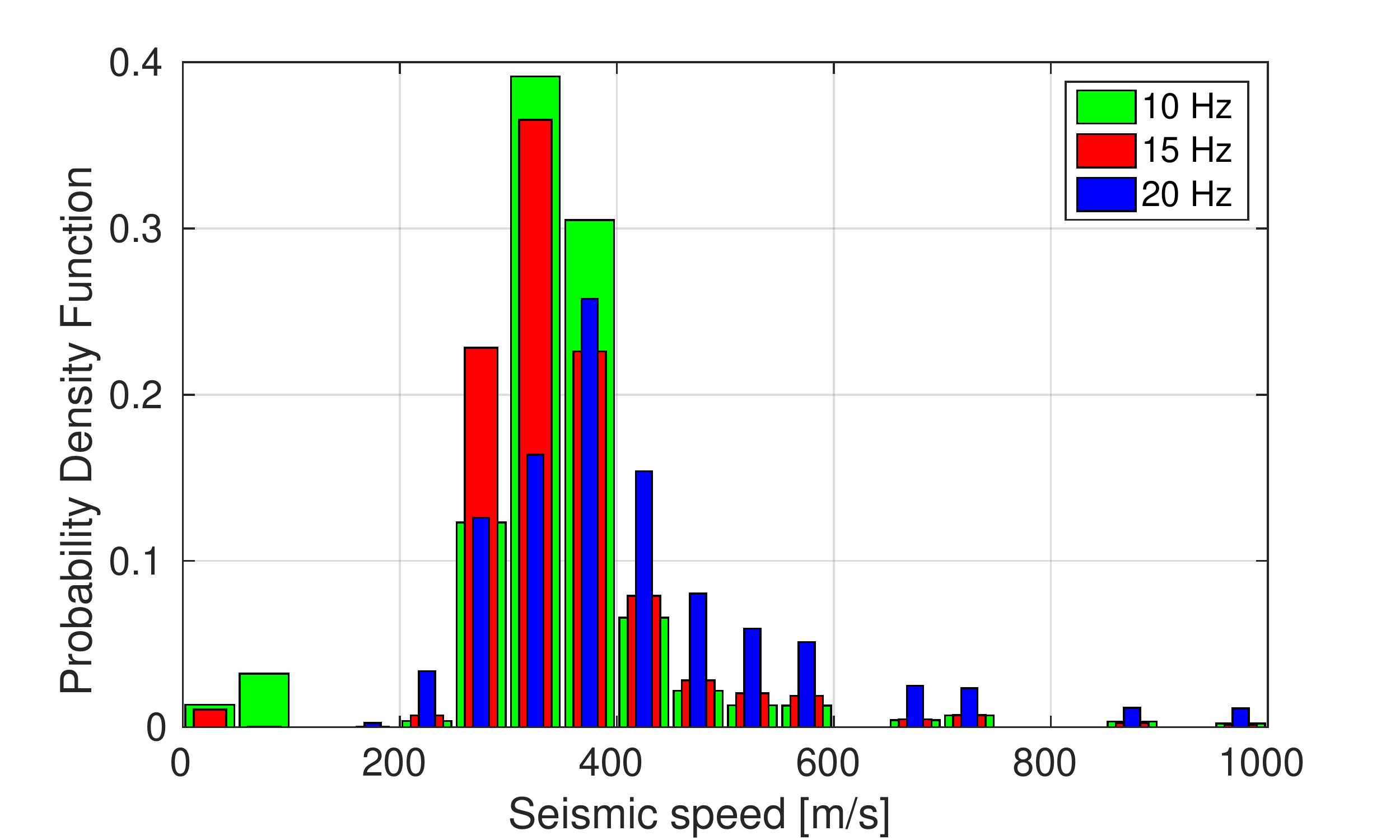}
 \caption{Histograms of seismic speed measurements using LHO array data at 10\,Hz, 15\,Hz, and 20\,Hz.}
 \label{fig:speeds}
\end{figure}

Beginning of October 2016, an array of 30 L-4C vertical-axis seismometers \footnote{www.sercel.com/products/Pages/seismometers.aspx} were deployed at the corner station building at LIGO Hanford. Concurrently, a single-axis tiltmeter was installed at the center of the array \cite{VeHa2014,HaVe2016}. The left of Figure \ref{fig:array} shows the locations of the seismometers in the vicinity of the vacuum enclosure. The data from the seismometers are conditioned, acquired digitally and saved at a 512\,Hz sampling rate. The configuration of seismometers was an approximately equidistant placement of a few meters between neighbors in the central part of the array and increased spacing along the edges. For the analysis, we divide the strain time-series data into 50\% overlapping 128\,s segments that are Hann$\mbox{-}$windowed. The first metric presented are percentiles of the PSDs, which show variations in the seismic field over the frequency band of interest. Each sensor in the map of Figure \ref{fig:array} is represented by 3 overlaid circles, the lower one representing the 10th percentile at 15\,Hz, the middle one the 50th, and the upper one the 90th percentile. Maps at other frequencies can be found in the Supplement. These maps allow us to identify local sources, or locations of vibration amplification due to interaction with infrastructure. The plot on the right shows the full spectra of all sensors for the 50th and 90th percentiles. The spectra have significant structure over the 10-20\,Hz band indicating the presence of several relatively narrow-band local sources.

The second metric presented is the complex coherence, $\gamma(f)$, between the seismometers in the array, calculated as $\gamma(f) = \frac{\langle x(f)\,y^*(f)\rangle}{\sqrt{\langle |x(f)|^2\rangle\langle |y(f)|^2\rangle\phantom{\big]}}}$
where $x(f)$ and $y(f)$ are the values of the Fourier Transform at a particular frequency $f$ for two seismometers.
This quantity allows for the characterization of the seismic field and is also an important quantity for the calculation of noise-cancellation filters \cite{Cel2000,CoMu2016}. Figure \ref{fig:coh} shows a measurement of $\Re(\gamma)$ between all 27 seismometers used for this study at 15\,Hz. Each coherence value is drawn at a coordinate, which corresponds to the relative position vector between the two sensors. Although there are some instances of inhomogeneities where high coherence points are near to low coherence points, in general the coherence evolves smoothly, and consistently with a Rayleigh-wave field. Homogeneity is a prerequisite for realizing NN cancellation with a relatively small number of seismometers (not more than 10 seismometers per test mass). The Supplement explores the variability in this quantity.
\begin{figure}[t]
 \includegraphics[width=\columnwidth]{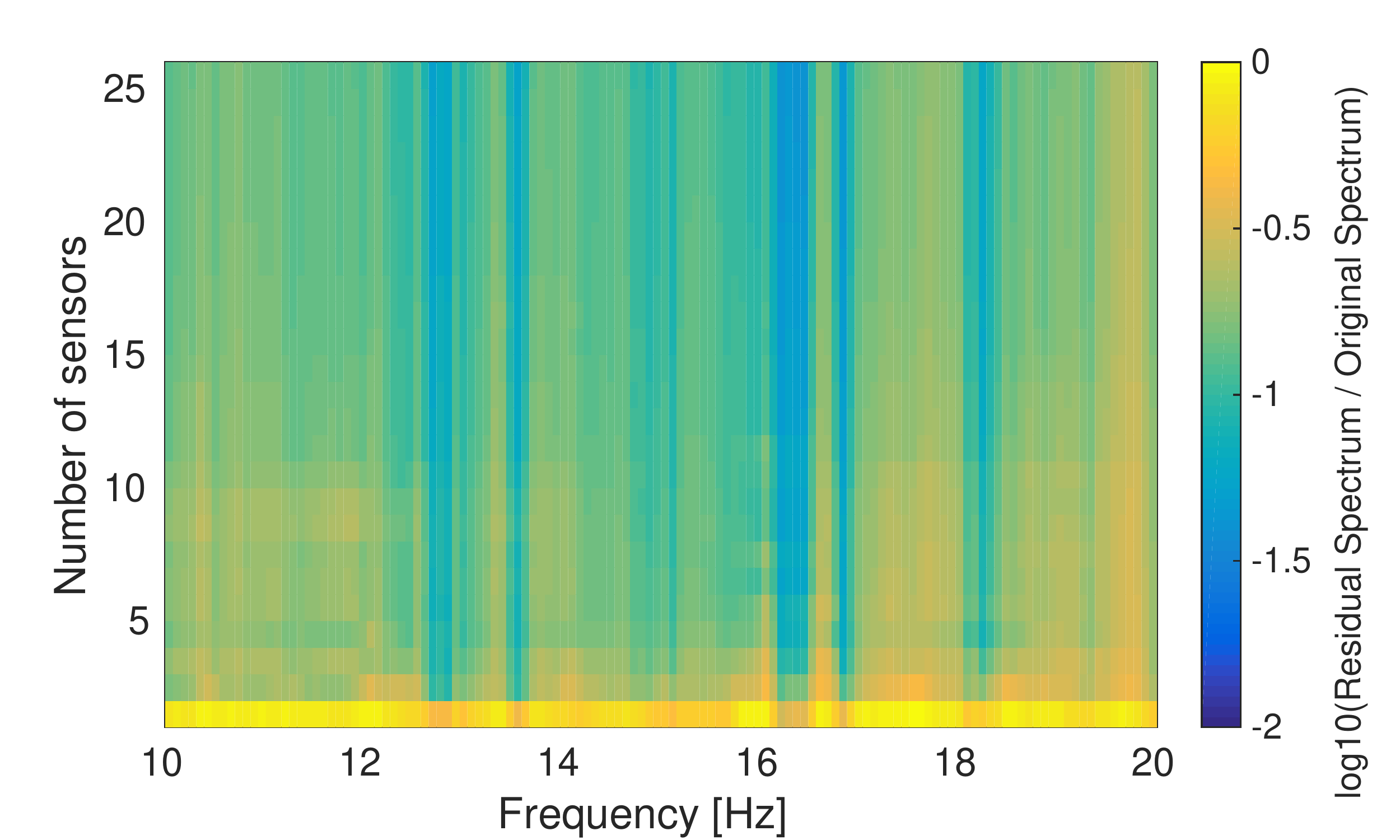}
 \caption{Expected residuals in the tiltmeter based on Equation (\ref{eq:R}) for all possible numbers of seismometers picked from the array such that residuals at 15\,Hz are minimized.}
 \label{fig:sub}
\end{figure}
Accounting for inhomogeneity is currently an open problem, and a much denser spatial sampling of the seismic field might be required to predict its effect on NN cancellation \cite{CoMu2016}.

The third metric presented is a measurement of wave speeds in the frequency range 10\,Hz -- 20\,Hz shown in Figure \ref{fig:speeds}, which is useful for further characterization of the field and predictions of the levels of Wiener filter subtraction that could be expected. We used the method of section 3.6.3 of \cite{Har2015}. The idea is to decompose the seismic field into plane harmonics and collect the phase speeds associated with the maximum-amplitude component. Consistent with measurements at the end stations \cite{CoMu2016}, the average velocities are about 300\,m/s at 10\,Hz and 15\,Hz and 380\,m/s at 20\,Hz, which is thought to be due to the concrete slab of the laboratory building, which has greater effect at shorter seismic wavelengths. Outliers in the histogram might be from body waves at higher speeds, or simply be the result of aliasing effects, which can happen when several waves at the same frequency simultaneously propagate through the array. Otherwise, these measurements are consistent with Rayleigh waves, which simplifies the NN modeling \cite{Har2015}. The width of the distribution can be explained by broadening due to sources being relatively close to the array giving rise to circular wave fronts, due to anisotropy of the ground, and potentially also due to seismic scattering (the latter two are likely minor effects at the LIGO sites since the soil does not vary significantly in horizontal directions over the extent of the array, and the surface is flat).

As discussed above, Wiener filters use correlations between reference data streams and a target data stream to give an estimate of the noise contributions to the target sensor present in the reference data streams as well \cite{BSH2008}. In the regime where all data streams have stationary noise, they are known to be optimal filters. In this work, we take a tiltmeter as our target sensor, with seismometers as reference data streams.

In the following, we will denote the cross-spectral densities between the $N$ seismometers in the array as $C_{\rm SS}(f)$, which is a $N\times N$ matrix. Similarly, we take the cross-spectral density between array and tiltmeter, which has $N$ terms, as $\vec C_{\rm ST}(f)$. We finally denote the PSD of the target sensor as $C_{\rm TT}(f)$. The Wiener filter is then constructed as $\vec w(f)=\vec C_{\rm ST}^\top(f)\cdot C_{\rm SS}^{-1}(f)$, where $C_{\rm ST}^\top(f)$ is the transpose of $\vec C_{\rm ST}(f)$ and $C_{\rm SS}^{-1}(f)$ is the inverse matrix of $C_{\rm SS}(f)$. The estimate of the target sensor data is then simply $\vec w(f) \cdot {\vec s}(f)$, where ${\vec s}(f)$ are the Fourier transforms at frequency $f$ of segments of data from all seismometers. Note that a similar definition of the Wiener filter can be given in time domain, realized for example as finite impulse-response filter.

\begin{figure*}[ht!]
 \includegraphics[width=3.5in]{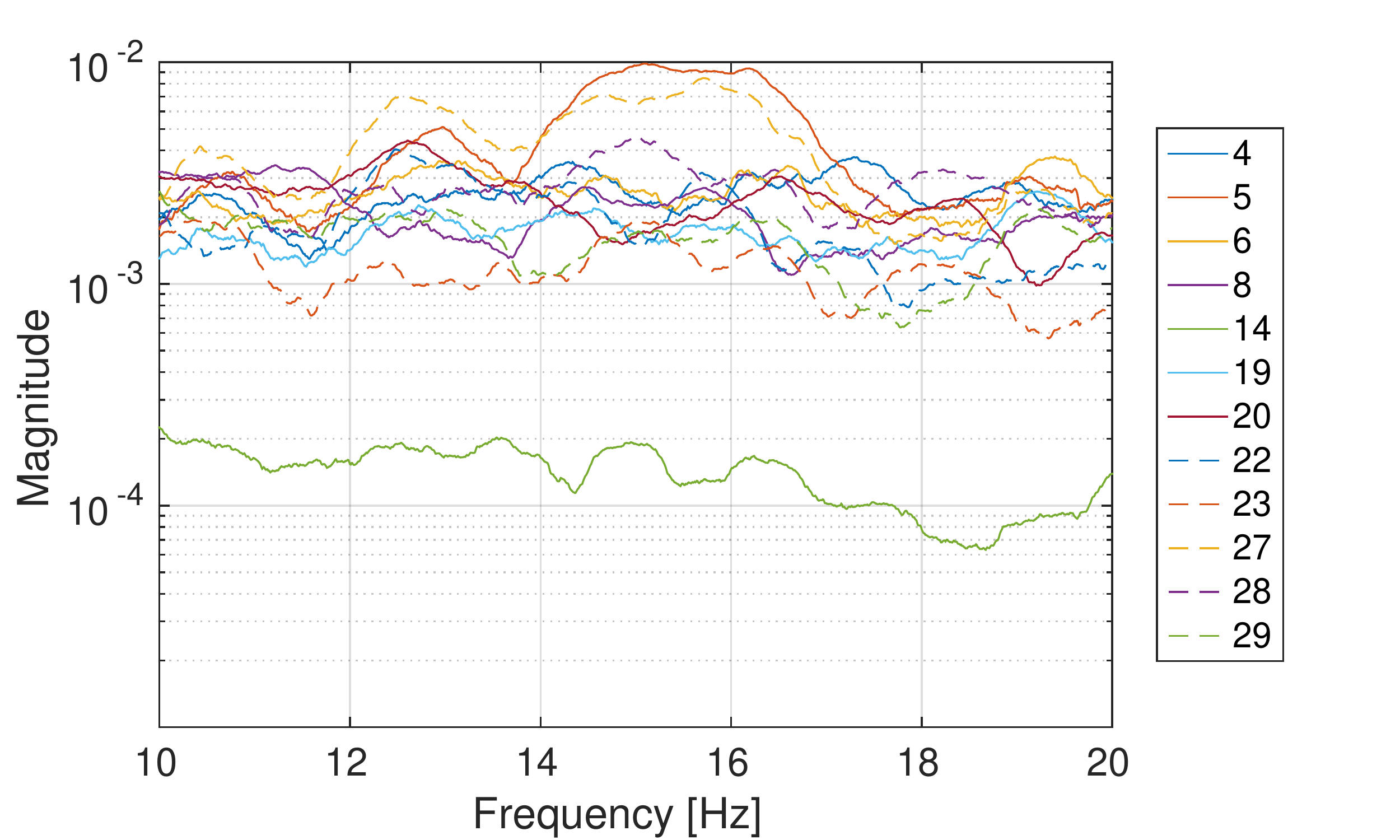}
 \includegraphics[width=3.5in]{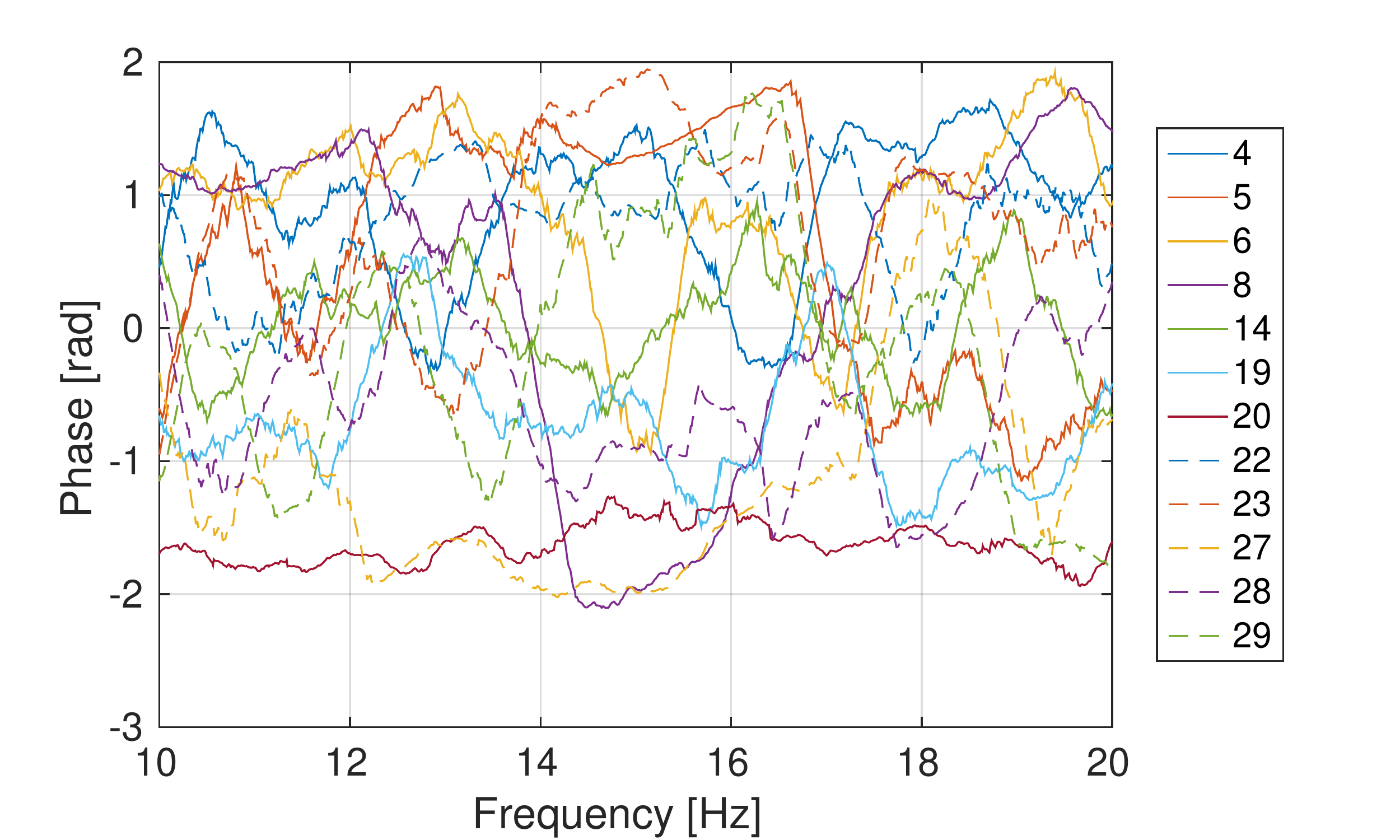}
 \caption{The magnitude (left) and phase (right) of the Wiener filter using 12 optimally picked witness channels.}
 \label{fig:bode}
\end{figure*} 

We can use the coherence results to determine the expected residuals. Based on the above quantities, the average relative noise residual $R$ is determined by
\begin{equation}
R = 1 - \frac{\vec{C}_{\rm ST}^\top \cdot C_{\rm SS}^{-1} \cdot \vec{C}_{\rm ST}}{C_{\rm TT}}.
\label{eq:R}
\end{equation}
In Figure \ref{fig:sub}, we use Equation (\ref{eq:R}) to determine the expected residuals for array configurations with optimal selection of a varying number of sensors. To determine the optimal configurations, we simply loop over all configurations to determine the configuration with the smallest residuals for a given number of seismometers picked from the array. Based upon this, the expected residuals rapidly converge after only a few seismometers. We will note that this is a different method than in Coughlin et al.~\cite{CoMu2016} where the sensor locations were allowed to vary arbitrarily and correlations were based on a model fit to the observed correlations. We show representative optimal array configurations in the Supplement. Based upon this, we expect significant suppression of tilt signals. When calculating Wiener filters from 30\,min of disturbance free data and then applying to the rest of the data, we achieve better than factor 10 reduction of tilt signals at some frequencies (see Supplement). 

The Wiener filter can be studied further by calculating its Bode plot shown in Figure~\ref{fig:bode}, which consists of the magnitude and phase of the filter for each witness channel as a function of frequency. The plots show that some seismometers form tiltmeter type configurations. These can be identified by searching for pairs of seismometers whose magnitudes are similar over the entire frequency band, and with a relative phase of about $180^\circ$. Sensors 4, 5, 20, and 27 form two such pairs. Many sensors contribute to the Wiener filter with similar magnitude. Combined with the observation from Figure \ref{fig:sub} that around 5 sensors or less are required to achieve most of the noise cancellation, similar magnitudes of sensors in a larger array means that the main impact of additional sensors is to average incoherent noise. We cannot fully explain the low magnitude of sensor 14 over the entire band. As the PSD maps indicate, see Figure \ref{fig:array} and supplement, the seismic signal at many frequencies is much stronger at sensor 14 than at other sensors used in Figure \ref{fig:bode}. This would explain the relatively low magnitude. However, its signal is weaker closer to 20\,Hz than in other sensors. This begs the question why the magnitude does not rise towards higher frequencies. It might be that its signal at higher frequencies is actually below its instrumental noise so that the Wiener filter suppresses injection of incoherent noise into the target channel. This hypothesis does not seem to be consistent though with such low magnitude values. It will be very important to study Bode plots of Wiener filters in greater detail to obtain an intuitive understanding of how the filter retrieves information from the witness channels, which could guide optimal placement of sensors even in inhomogeneous seismic fields where numerical methods still fail. 

\begin{figure}[ht!]
 \includegraphics[width=\columnwidth]{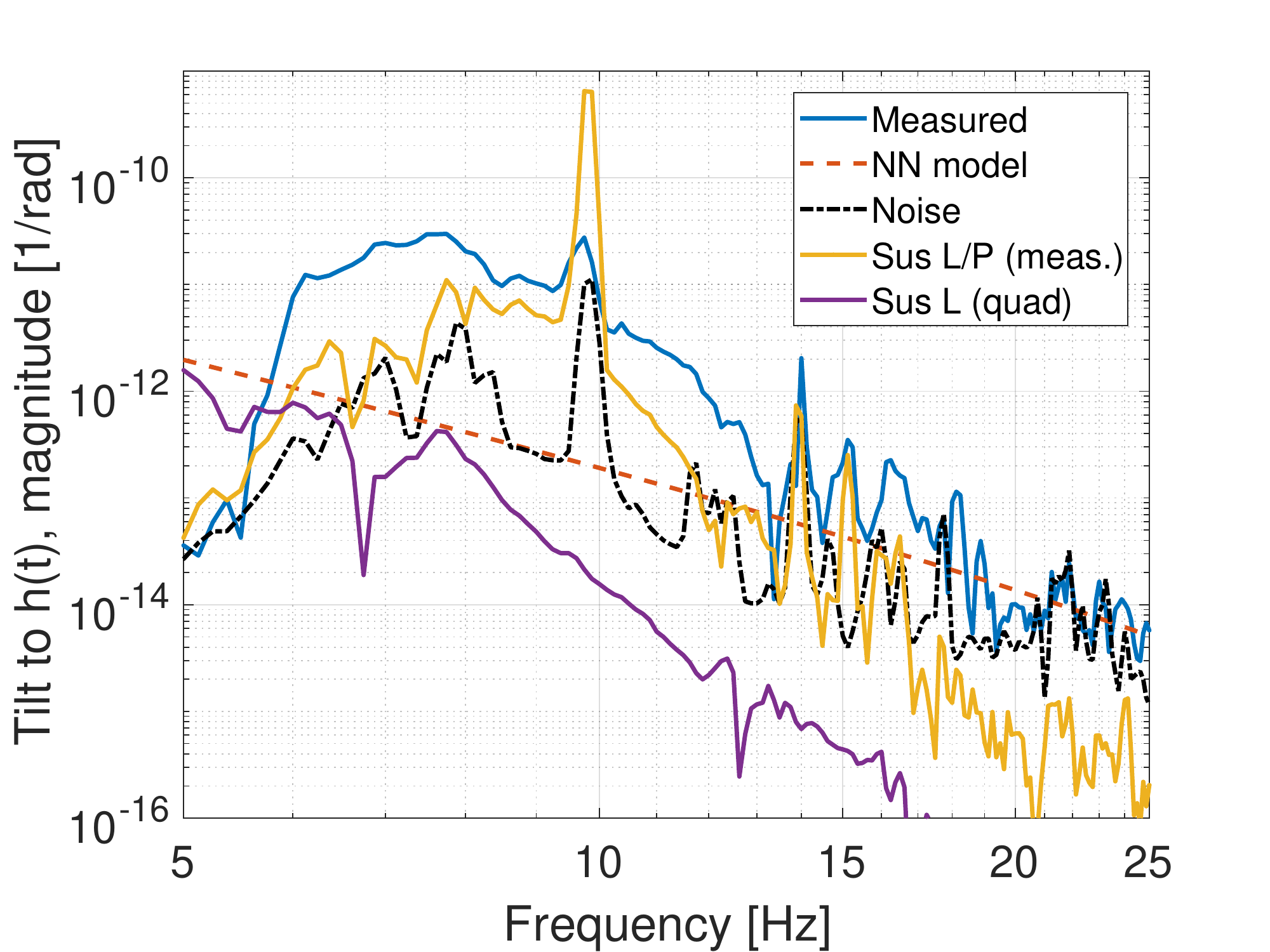}
 \caption{Ground tilt to GW data couplings.}
 \label{fig:couplings}
\end{figure}

Last, we present measurements of transfer functions between ground tilt and GW data 'h(t)' shown in Figure \ref{fig:couplings}. Newtonian noise is predicted to lie about a factor 100 below other instrumental noise during the second science run, so that long correlation times need to be used to observe gravitational coupling. Our measurements use about 1 month of data with the interferometer locked starting in December 2016. A NN model is plotted for an isotropic, homogeneous field taking into account that seismic waves generally produce NN through both test masses of the corner station, and considering the positions of the two test masses and tiltmeter. The direct measurement of the transfer function from ground tilt to h(t) is shown as blue, solid line. Additional couplings were investigated through the seismic isolation system: displacement of the test-mass suspension-point along the arm (Sus L), and pitch of the suspension table (Sus P), i.e., a rotation around the horizontal axis perpendicular to the arm. The measured transfer function from ground tilt to Sus L can be subsequently passed through a model of the quadruple suspension system (solid, violet line), or through a measured transfer function between Sus L and h(t) (solid, yellow line). The yellow line corresponds in fact to the sum of the two transfer functions through Sus L and P (which does not accurately represent the total coupling through Sus L/P since it ignores correlations between these two channels). Last, the measurement noise is shown, which was calculated by sliding the ground tilt and h(t) time series against each other by about 1000\,s, and also by an analytic Gaussian model of measurement noise (which both give almost identical curves).

With the exception of a few frequencies, the observed ground tilt to h(t) coupling (solid, blue) lies well above the measurement noise (dot-dashed, black) up to about 20\,Hz. It is inconsistent with contributions from suspension point motion, which by itself is inconsistent with a model of mechanical coupling through the suspension stages (hinting towards additional coupling mechanisms). The observed ground tilt to h(t) coupling is consistent with a simple (isotropic, homogeneous) NN model above about 13\,Hz where deviations between observation and model are small enough to be explained by anisotropies (and potentially inhomogeneities) of the seismic field. A detailed analysis of these effects and more careful studies of potential additional coupling mechanisms are under way.

In summary, we have used dedicated measurements at the LIGO Hanford site to predict NN cancellation levels. We showed how we were able to achieve significant subtraction in line with expectations based on correlation measurements. We used a tiltmeter near one of the test masses at the corner station of the LIGO Hanford detector as a proxy for NN in a gravitational-wave detector. We showed that significant subtraction of the tilt signal is achievable with only a few seismometers. We also found that it is unlikely to be necessary to update the Wiener filters often, since subtraction performance did not change significantly over the course of months. Observations of ground tilt to h(t) coupling suggest that NN detection might be feasible in the near future.

\acknowledgments
MC was supported by the David and Ellen Lee Postdoctoral Fellowship at the California Institute of Technology. NM acknowledges Council of Scientific and Industrial Research (CSIR), India for providing financial support as Senior Research Fellow. BS was supported by ARC Future Fellowship FT130100329 and DM and BS are supported by the ARC Centre of Excellence for Gravitational Wave Discovery. Thanks to the Seismic Working Group of the LIGO Scientific Collaboration for the suspension model. LIGO was constructed by the California Institute of Technology and Massachusetts Institute of Technology with funding from the National Science Foundation and operates under cooperative agreement PHY-0757058. The authors thank to the LIGO Scientific Collaboration for access to the data and gratefully acknowledge the support of the United States National Science Foundation (NSF) for the construction and operation of the LIGO Laboratory and Advanced LIGO as well as the Science and Technology Facilities Council (STFC) of the United Kingdom, and the Max-Planck-Society (MPS) for support of the construction of Advanced LIGO. Additional support for Advanced LIGO was provided by the Australian Research Council. This paper has been assigned LIGO document number LIGO-P1800049.

\section*{References}
\bibliographystyle{iopart-num}
\bibliography{references}

\appendix

\section{Power Spectral Densities}

\begin{figure*}[ht!]
\includegraphics[width=3.5in]{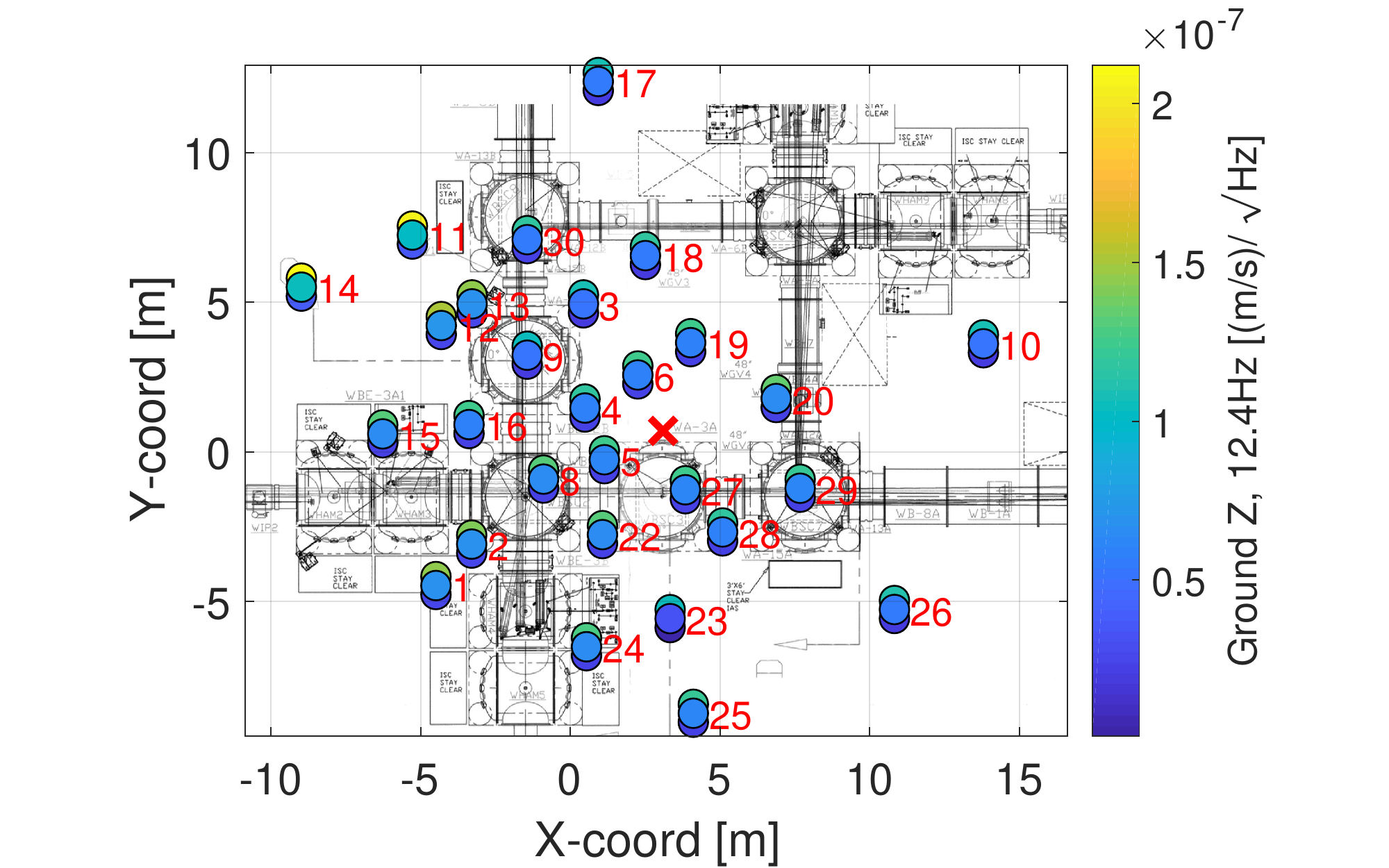}
\includegraphics[width=3.5in]{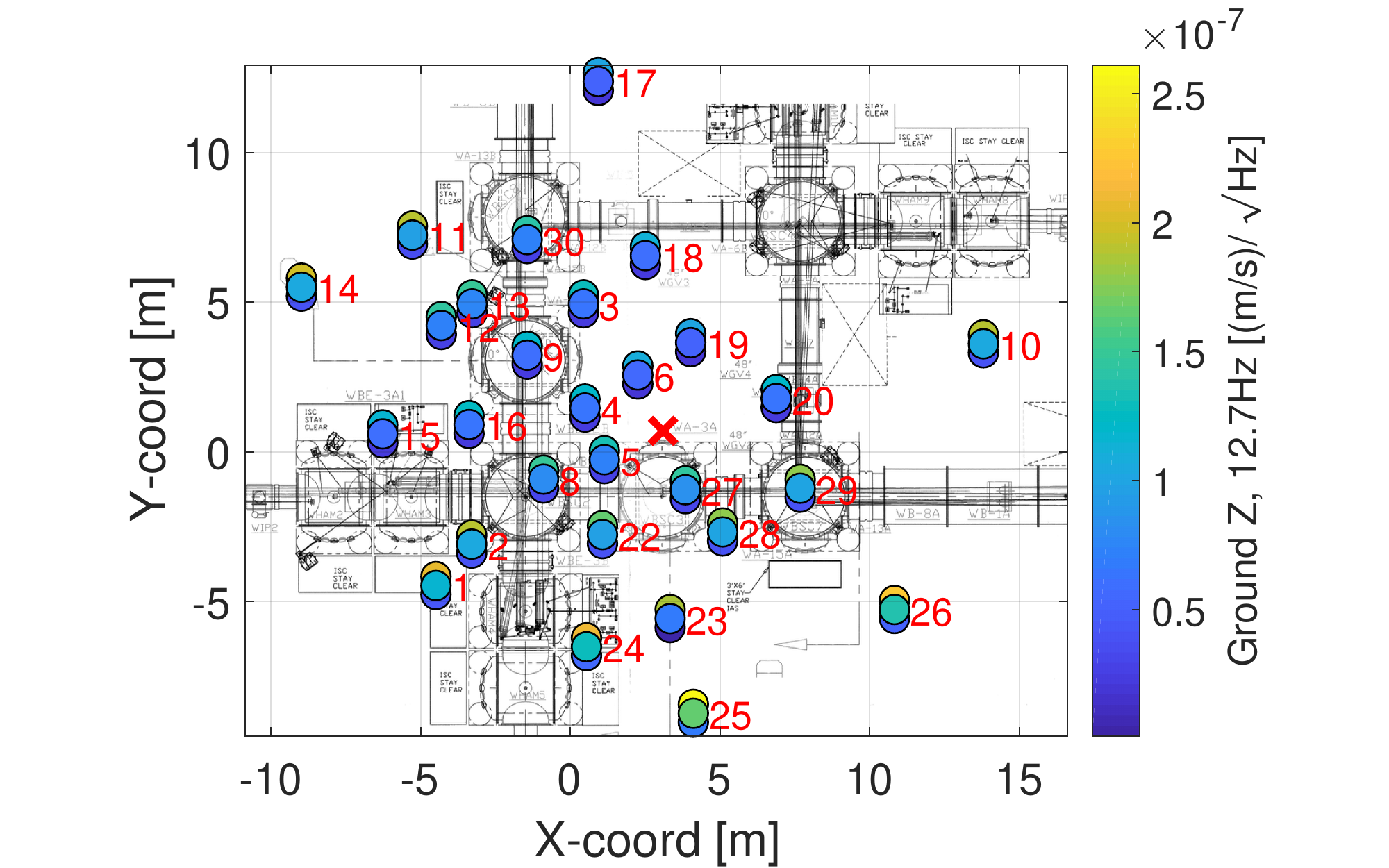}
\includegraphics[width=3.5in]{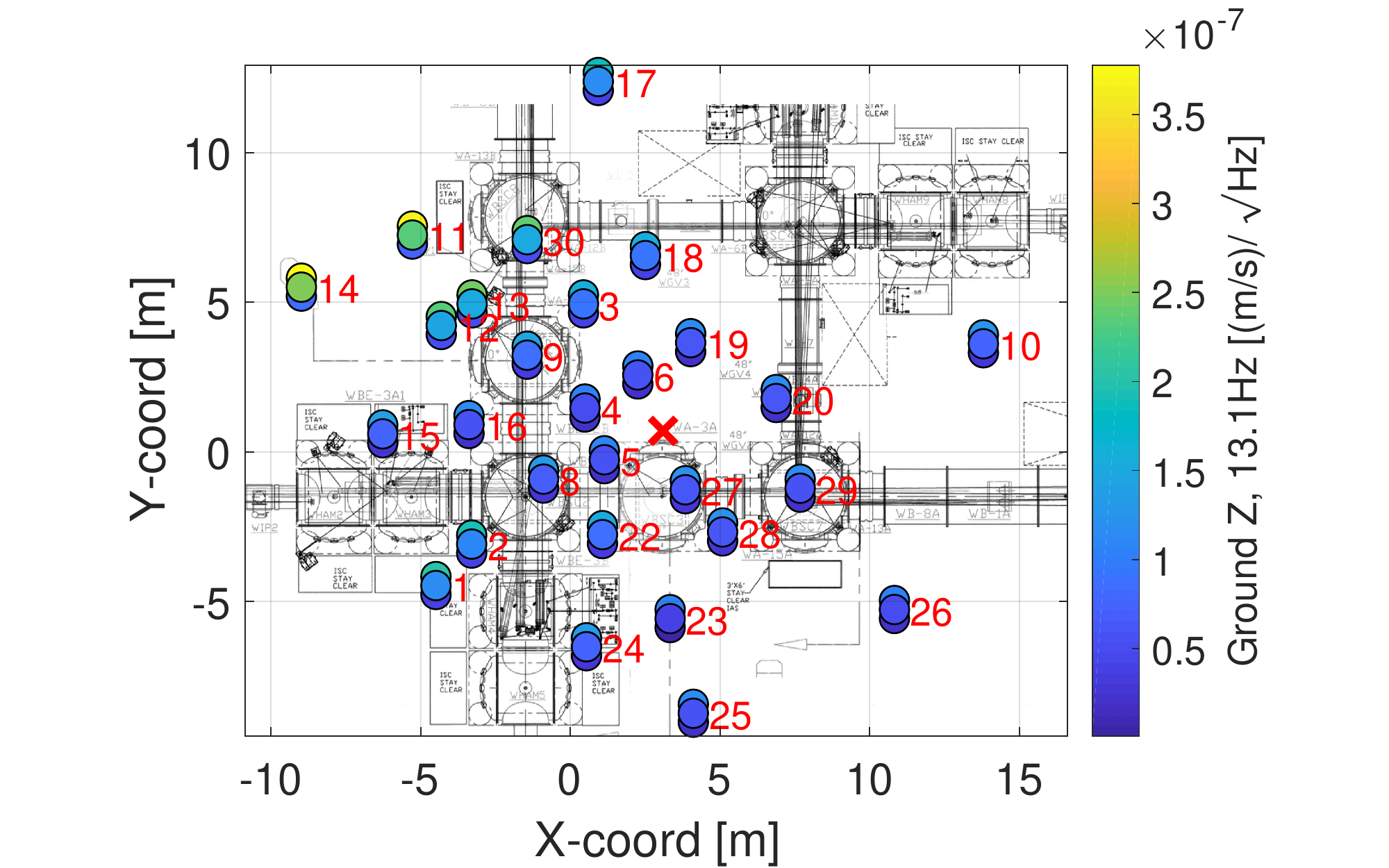}
\includegraphics[width=3.5in]{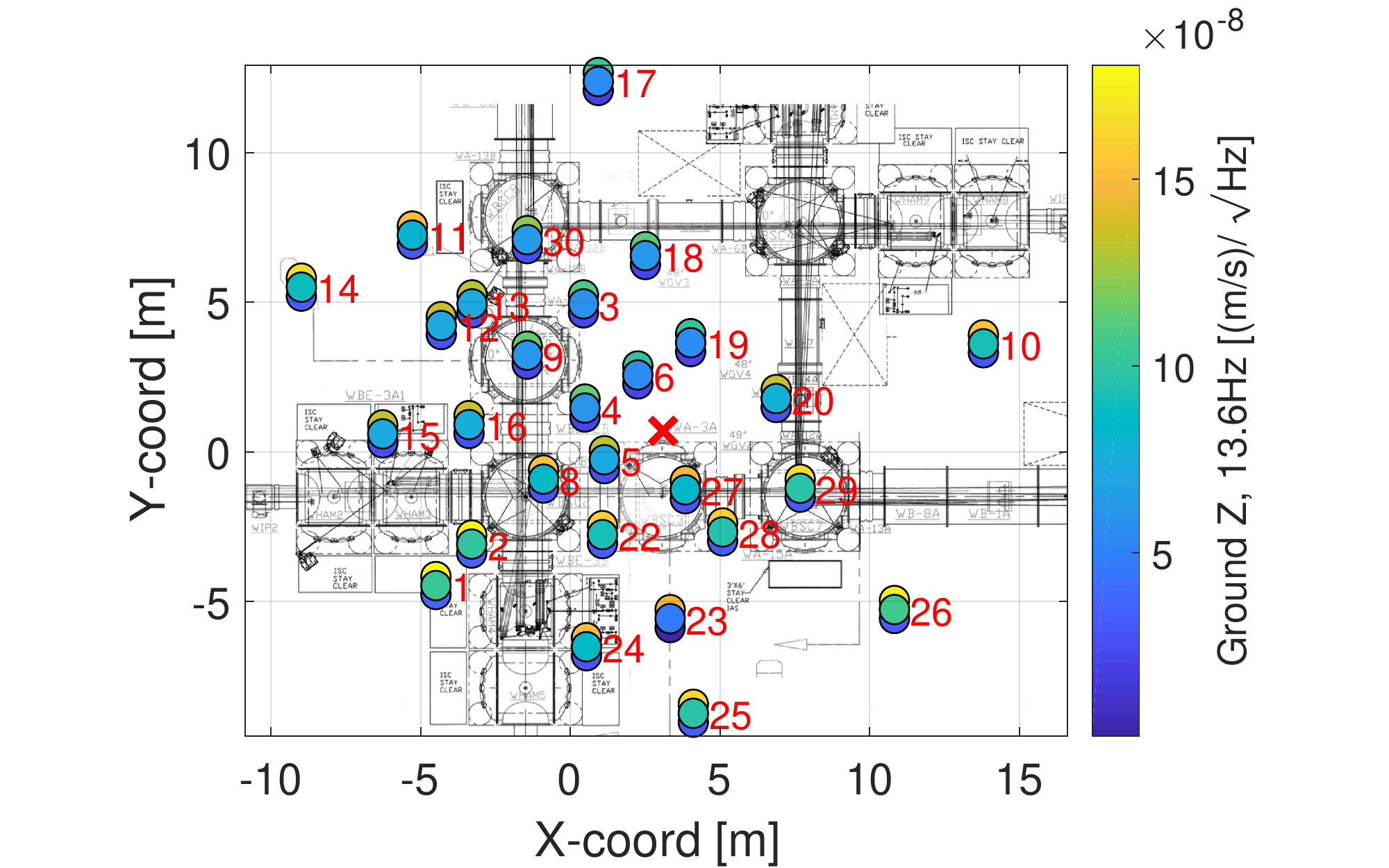}
\includegraphics[width=3.5in]{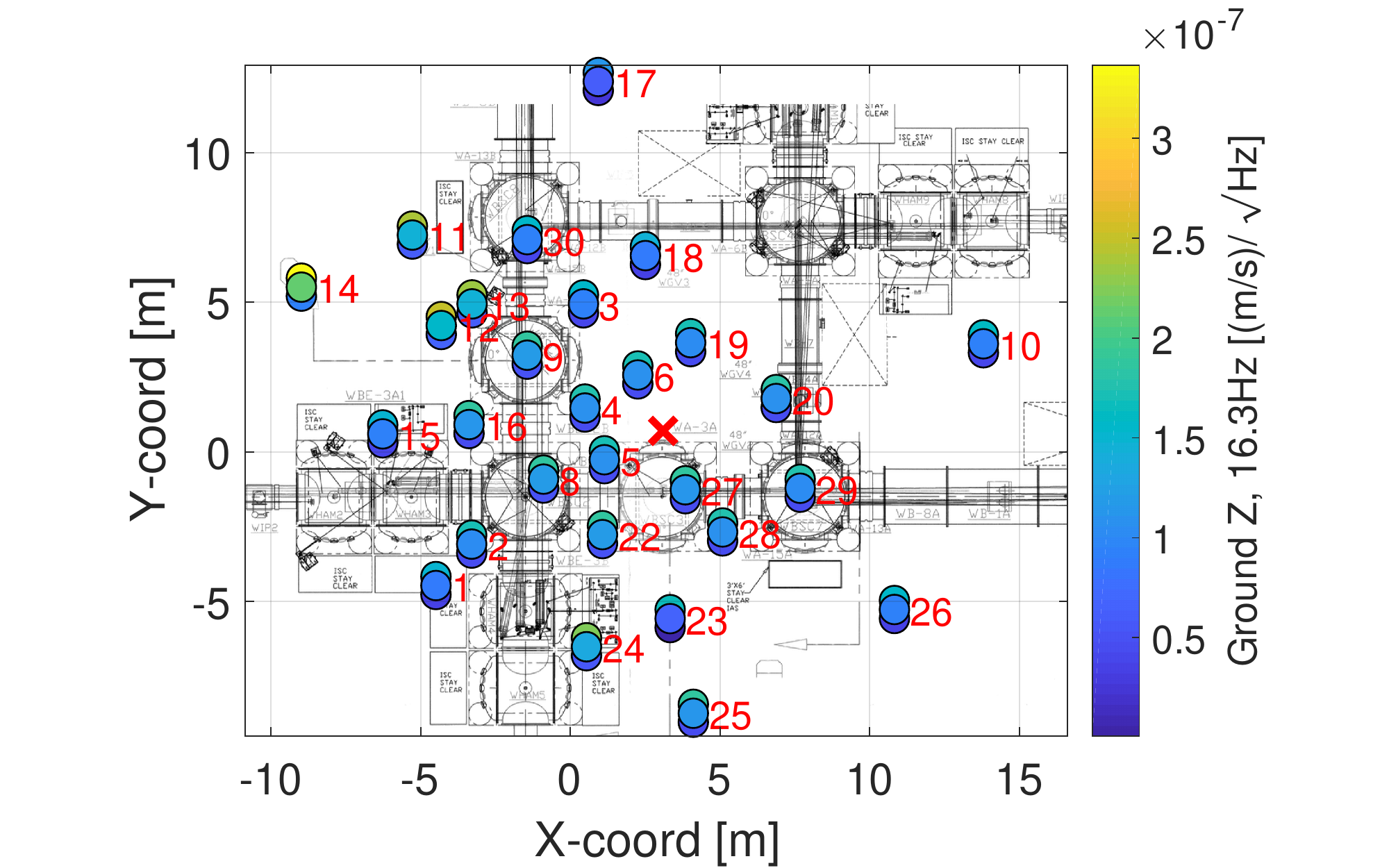}
\includegraphics[width=3.5in]{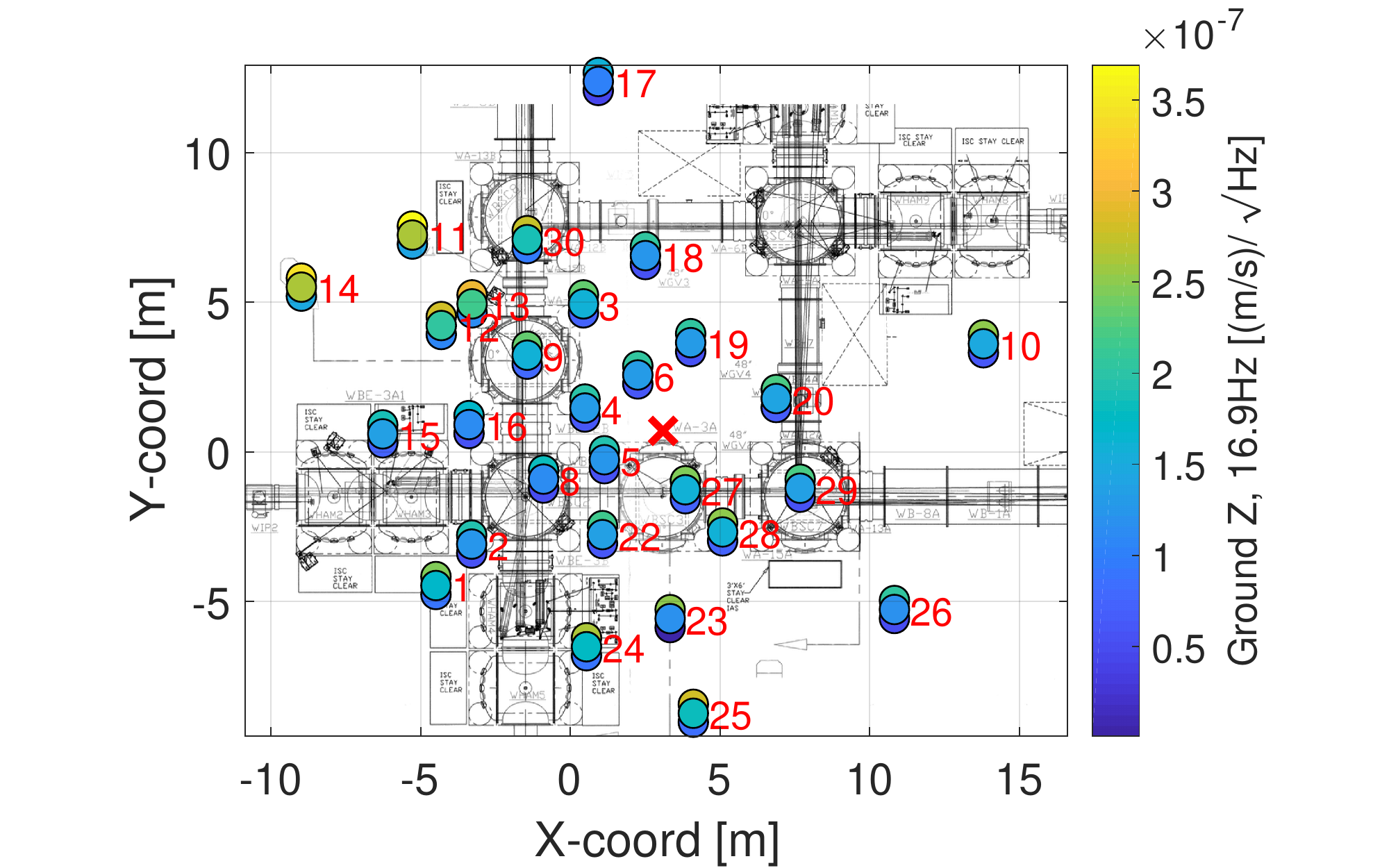}
\includegraphics[width=3.5in]{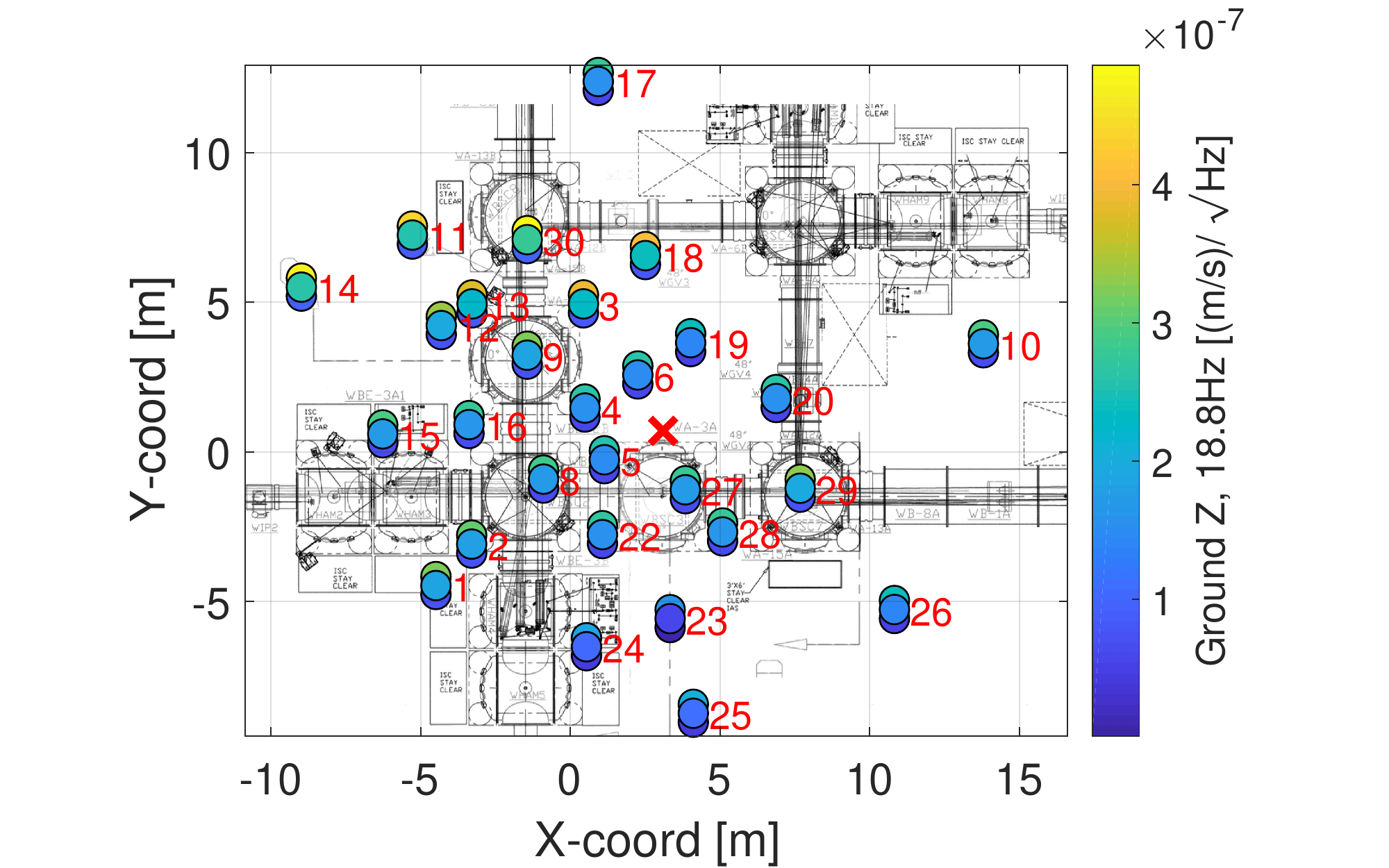}
\includegraphics[width=3.5in]{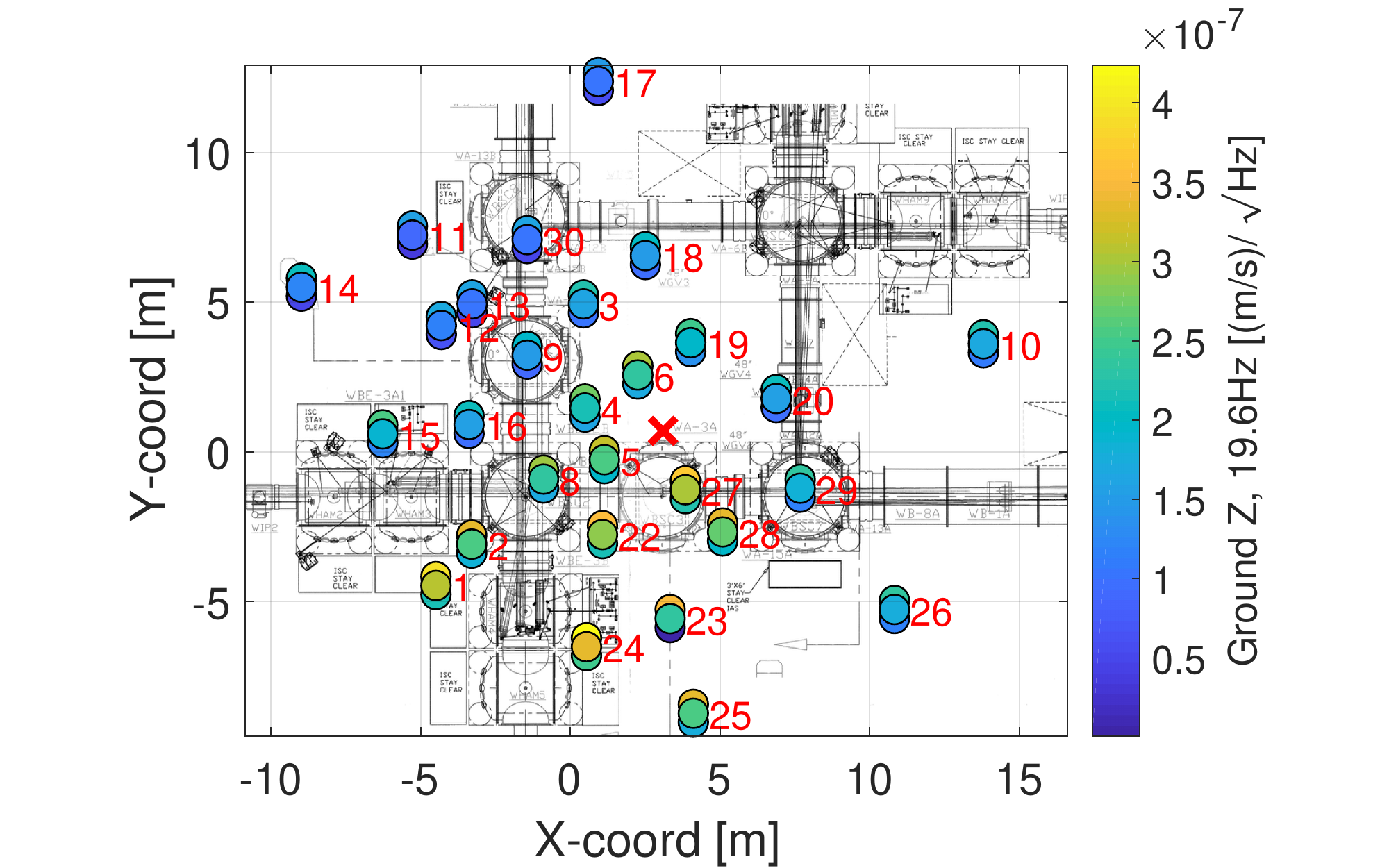}
\caption{Layout of the instrument floor at the corner station of the LIGO Hanford Observatory. The set of colored circles indicate placement of seismometers in the vicinity of the vacuum enclosure. At each set, the 10\%, 50\%, and 90\% percentile vertical power spectral density is indicated from bottom to top. The frequencies shown are 12.4, 12.7, 13.1, 13.6, 16.3, 16.9, 18.8, and 19.6\,Hz.}
\label{fig:arrayall}
\end{figure*}

Figure~\ref{fig:arrayall} shows the locations of the seismometers in the vicinity of the vacuum enclosure, as well as the 10\%, 50\%, and 90\% percentile vertical power spectral density at a variety of frequencies of interest.

\section{Variability in the correlation function}
\label{sec:cohstd}

\begin{figure}[t]
 \includegraphics[width=3.5in]{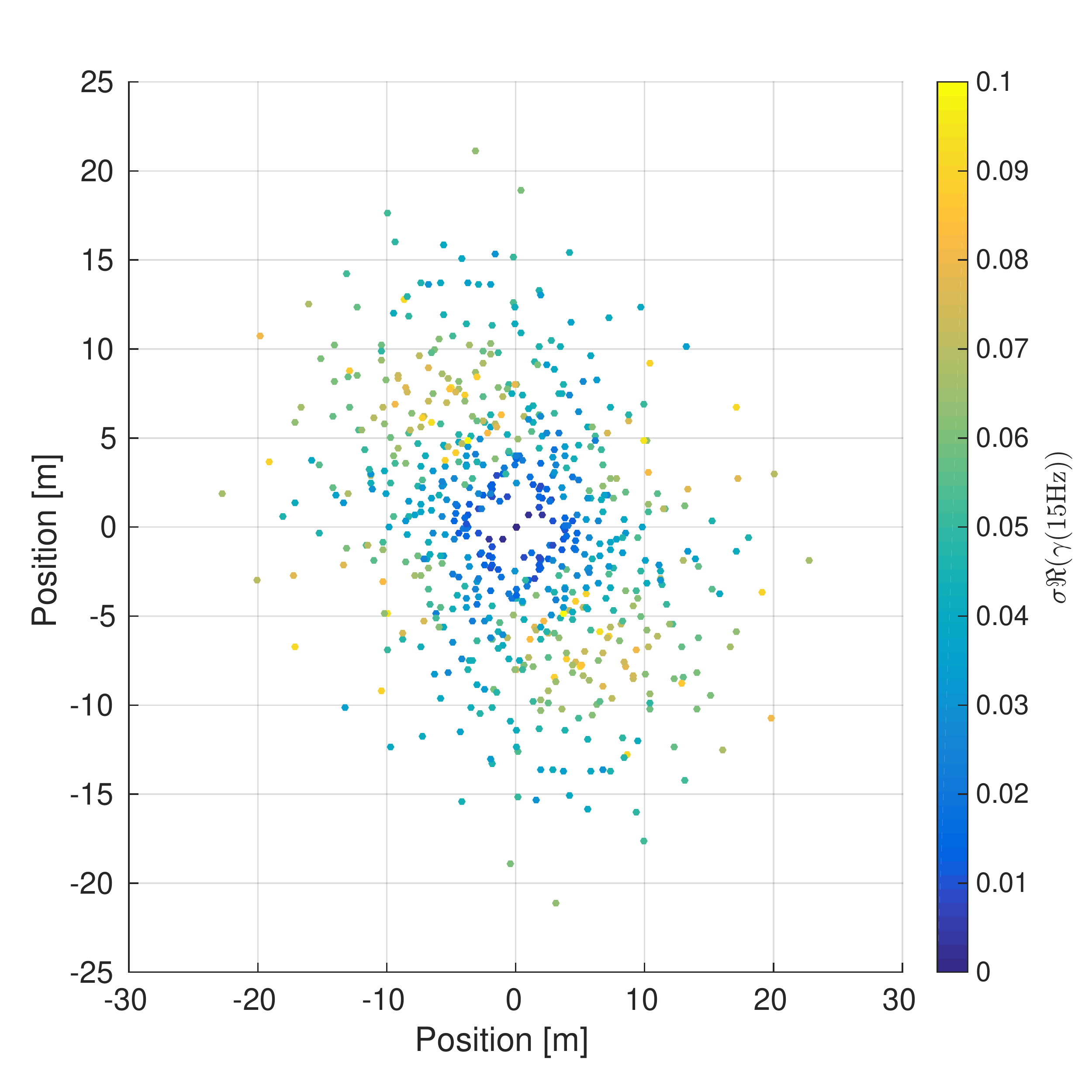} 
 \caption{
 Standard deviation in the measured correlation function for the LIGO Hanford corner station at 15\,Hz.}
 \label{fig:cohstd}
\end{figure}

Figure \ref{fig:cohstd} shows the standard deviation in the measurement of $\Re(\gamma)$ between all 27 seismometers used for this study at 15\,Hz. The coherences are calculated at a rate of once per day, and the standard deviation is computed from this distribution. As expected, the standard deviation is smallest near the center of the distribution, where both seismometers are most near to one another. The standard deviation is generally higher as the coherence becomes smaller.

\section{Optimal Array configurations}
\label{sec:optimal}
\begin{figure*}[ht!]
 \includegraphics[width=3.5in]{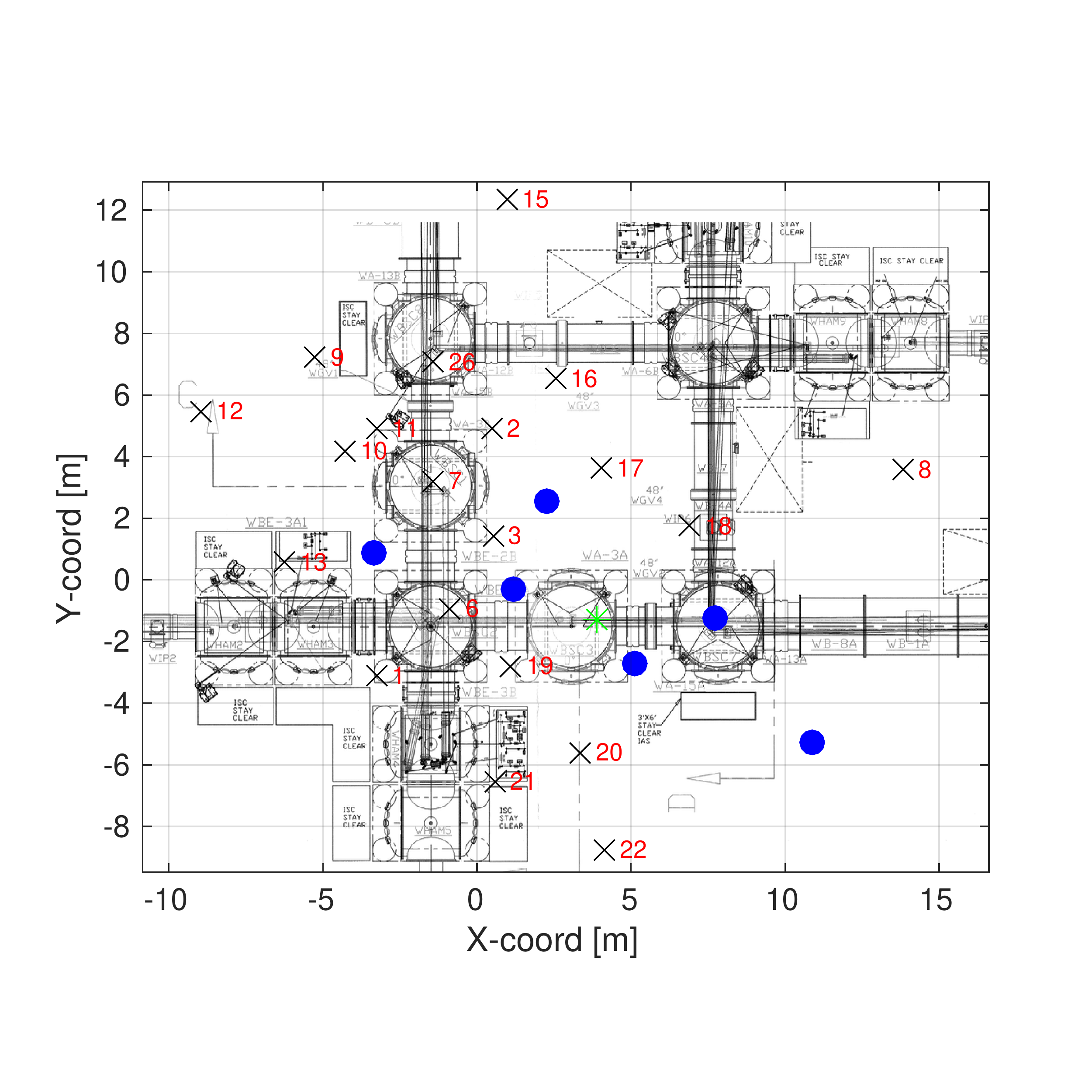}
 \includegraphics[width=3.5in]{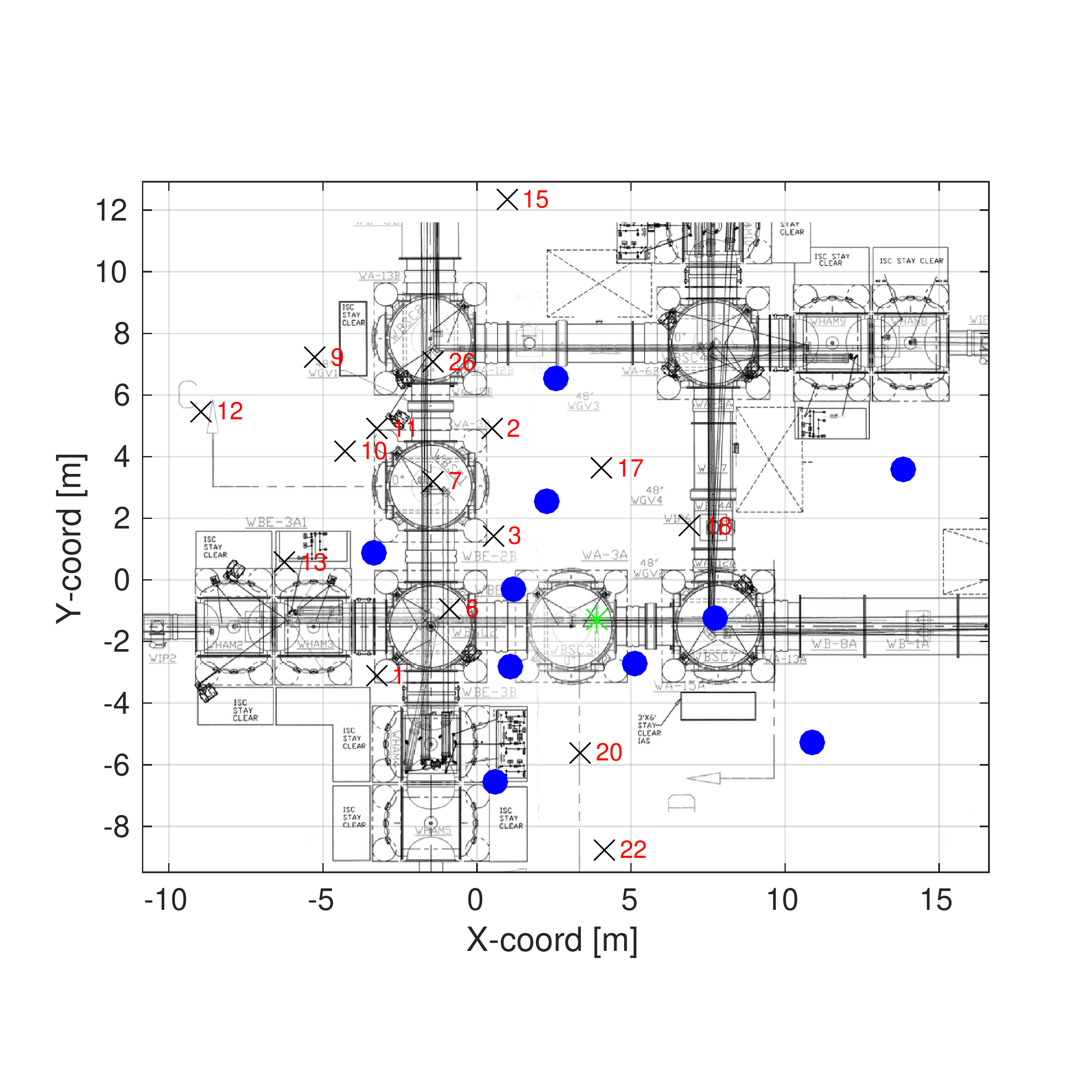}
 \includegraphics[width=3.5in]{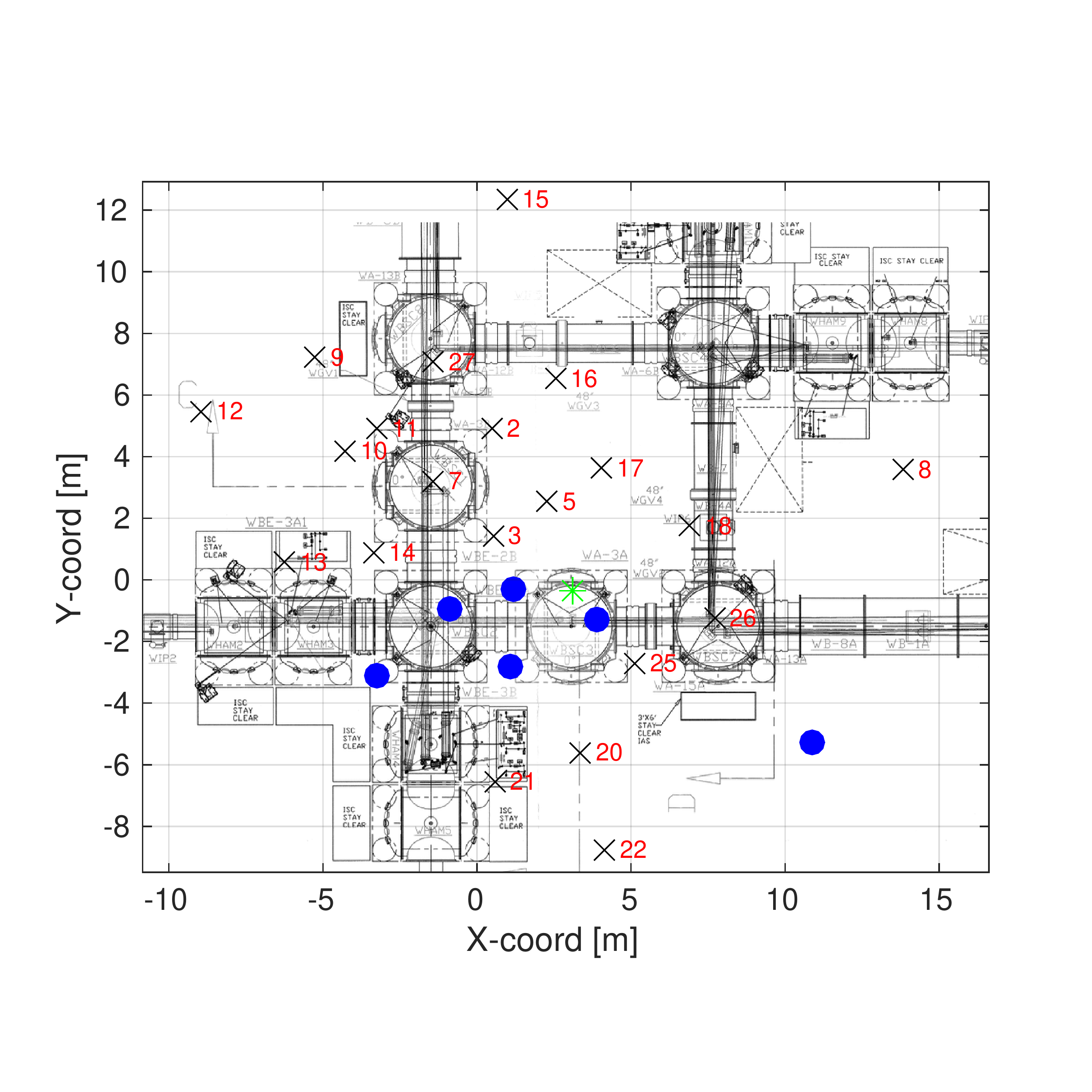}
 \includegraphics[width=3.5in]{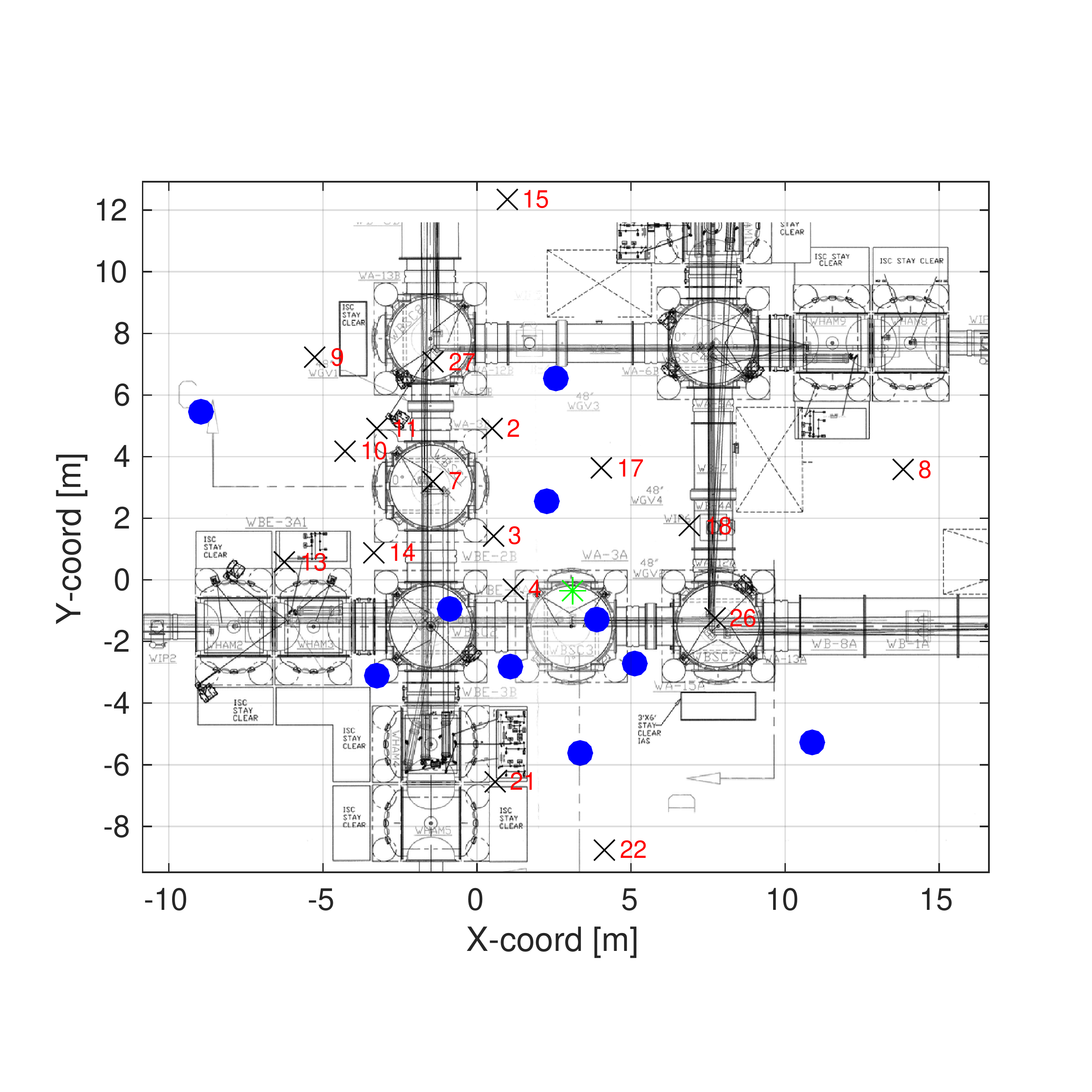}
 \caption{Optimal 6 and 10 sensor arrays (left and right columns) for an seismometer and tiltmeter (top and bottom rows). The large blue circles indicate the locations of the seismometers that are the optimal witness sensors, while the green star is the location of the target sensor (seismometer or tiltmeter).}
 \label{fig:bruteforce}
\end{figure*}

In figure~\ref{fig:bruteforce}, we show the optimal 6 and 10 sensor arrays (left and right columns) for an seismometer and tiltmeter (top and bottom rows).

\section{Subtraction with tiltmeter as target sensor}
Figure \ref{fig:sub} shows the spectrum of subtraction residuals of the tiltmeter relative to its average spectrum. The dashed curve is calculated using all seismometers as witness channels of the Wiener filter. The solid curve is calculated by using the one most effective seismometer at each frequency. The residual spectrum with all seismometers as witness channels is consistent with estimates of the tiltmeter instrumental noise (which is not accurately known though). 

\begin{figure}[ht!]
 \includegraphics[width=3.5in]{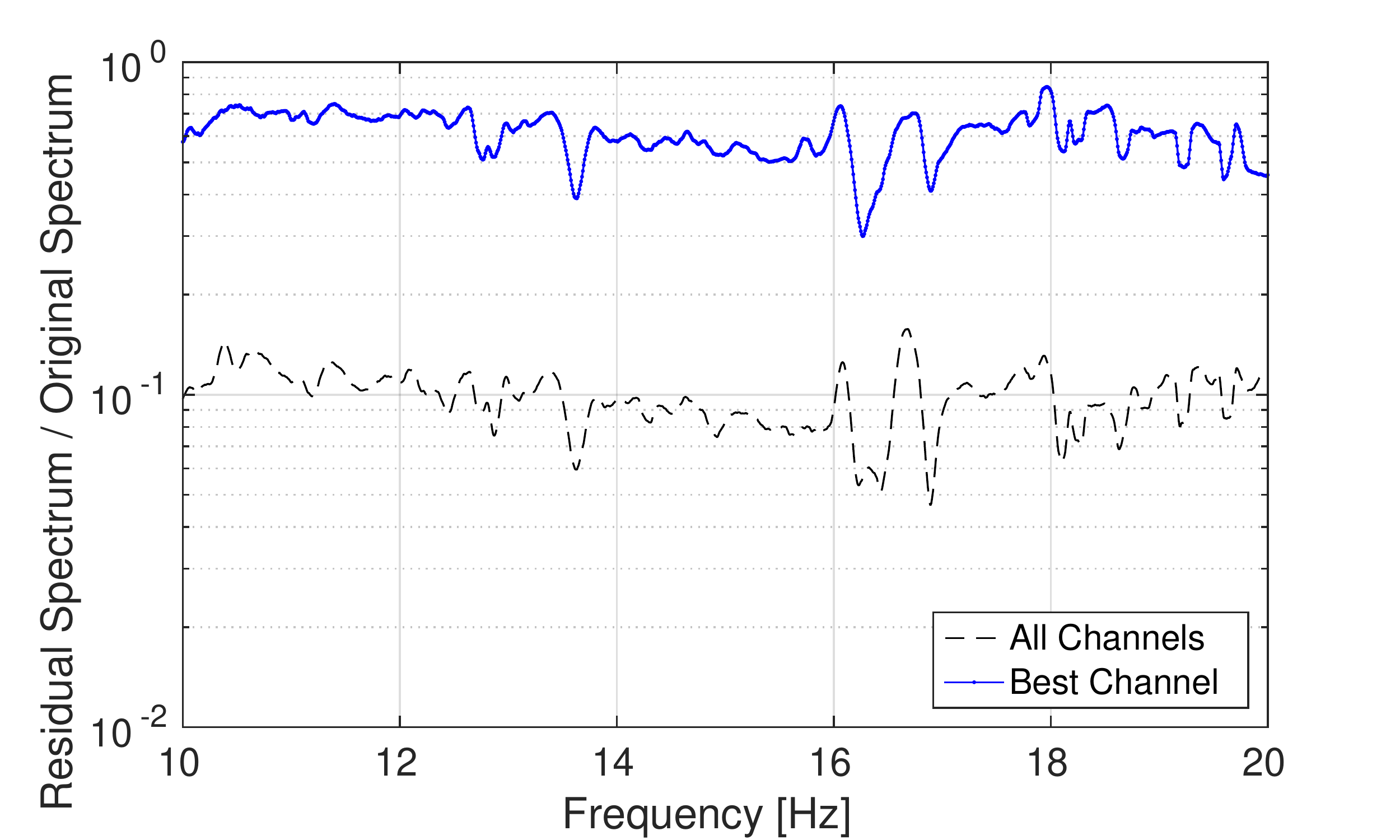}
 \caption{Ratio of the square-roots of PSDs of tiltmeter signal before and after Wiener filter subtraction using all available seismometers as witnesses or the single most effective seismometer picked for each frequency.}
 \label{fig:sub}
\end{figure}

\label{sec:seismometer}

\begin{figure*}[ht!]
 \includegraphics[width=3.5in]{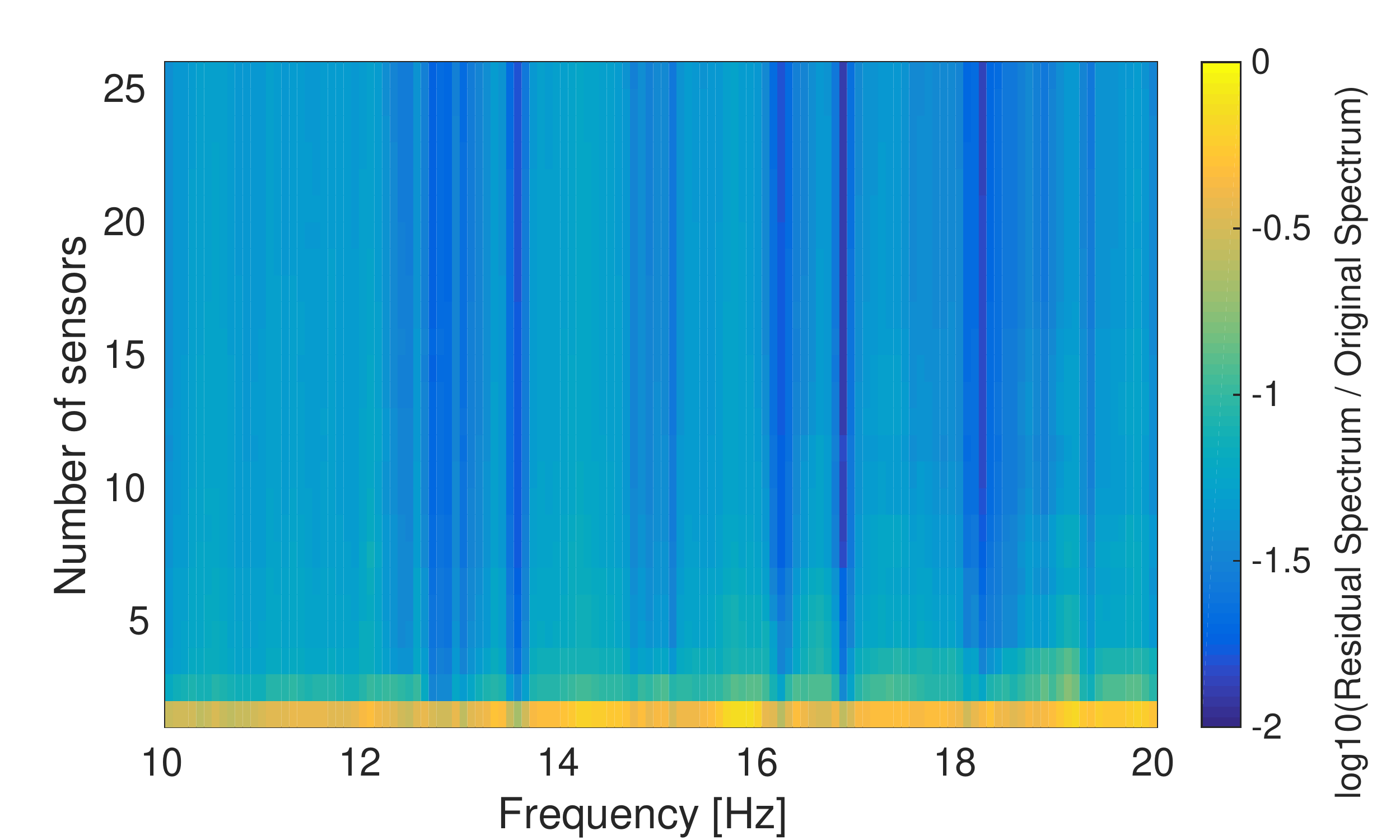}
 \includegraphics[width=3.5in]{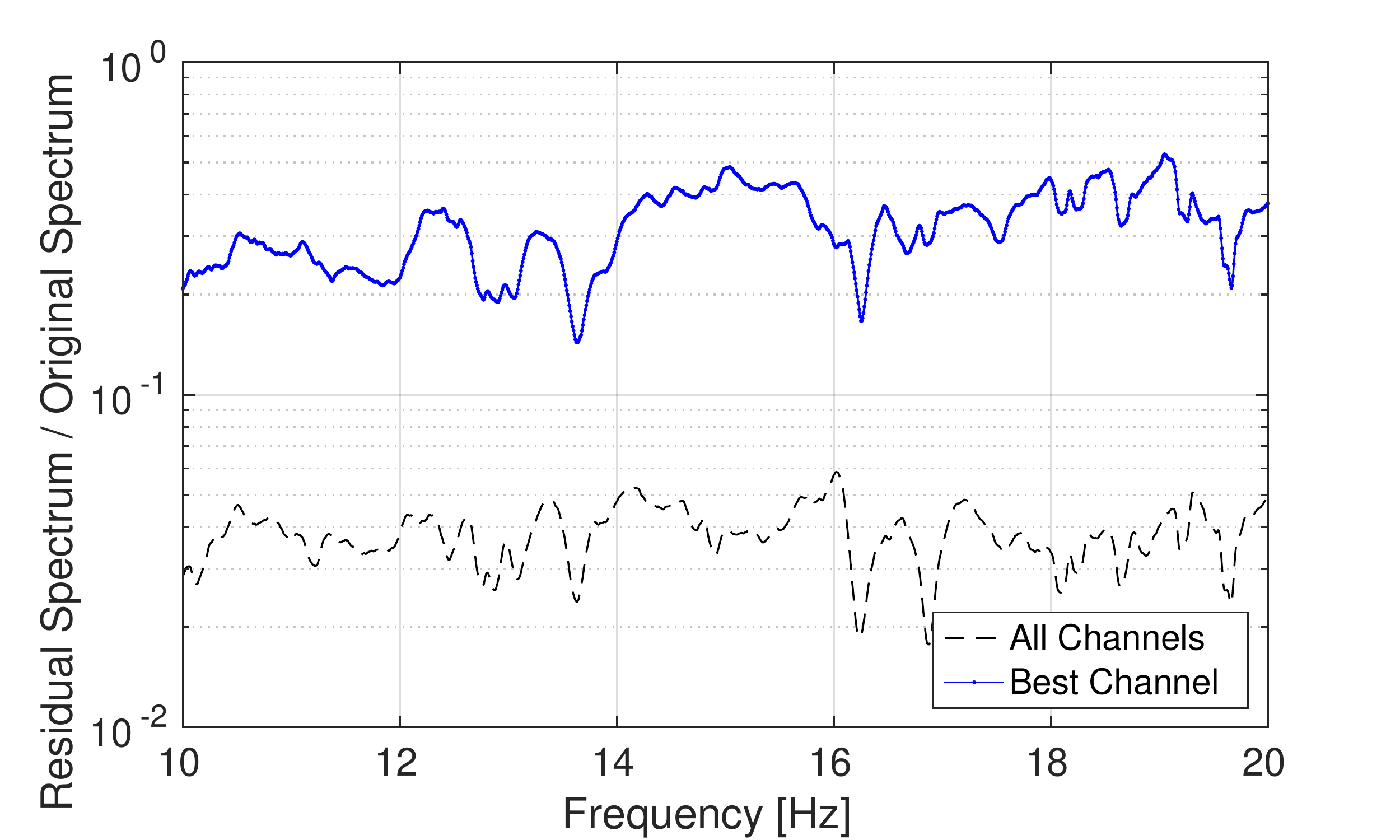}
 \caption{On the left is the expected residuals based on Equation (1) of the letter, using the center seismometer as target and using all other seismometers in the array as witnesses. On the right is the ratio of the auto power spectral density before and after Wiener filter subtraction.}
 \label{fig:subseismometer}
\end{figure*}

Similar to the discussion of the tiltmeter subtraction, on the left of figure~\ref{fig:subseismometer}, we show the subtraction using a frequency domain Wiener filter between 10-20\,Hz applied to an seismometer at the center of the array. The results are consistent with the expectations demonstrated in the top row of figure~\ref{fig:subseismometer}.

\section{Timescales for Wiener filter subtraction}
\label{sec:wiener}

\begin{figure*}[ht!]
 \includegraphics[width=3.5in]{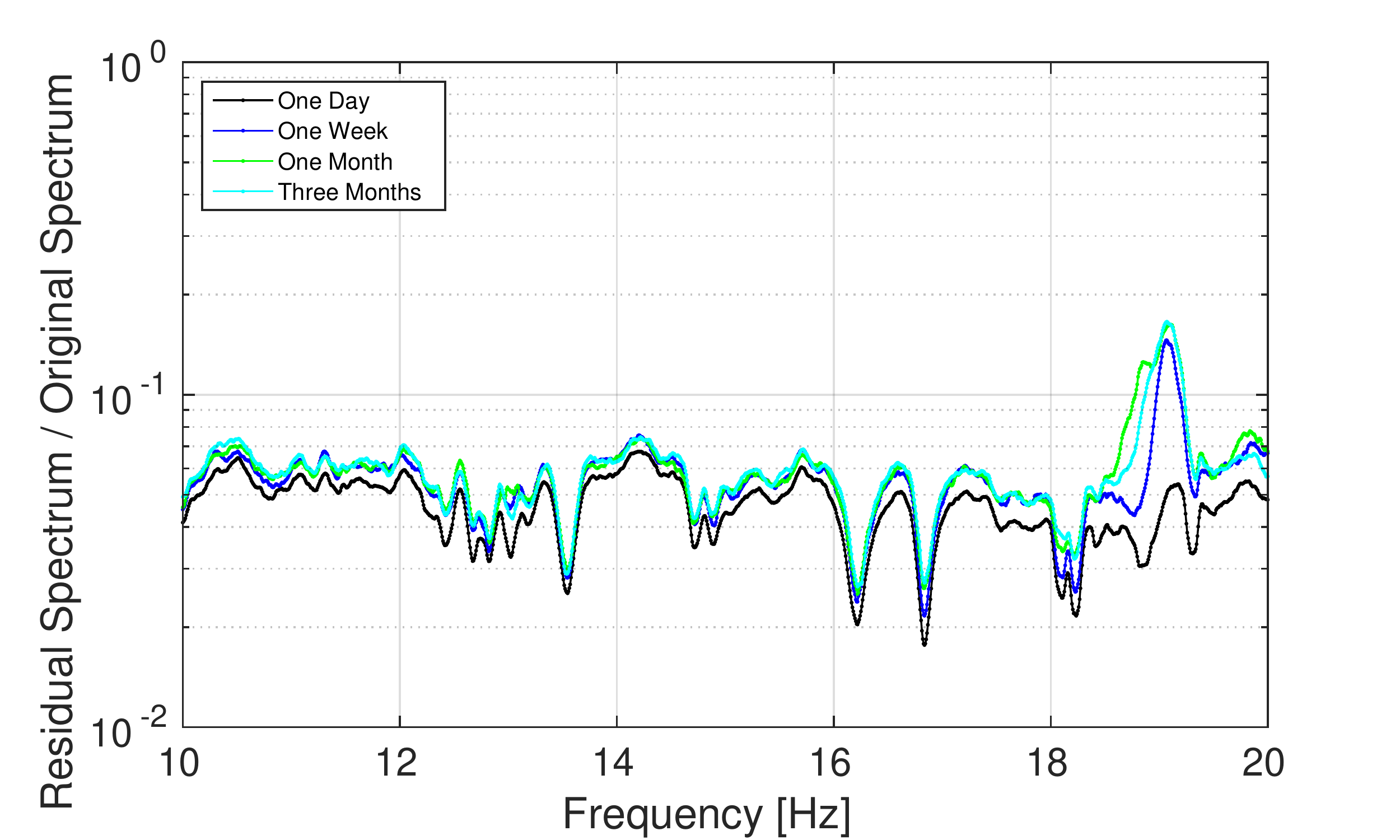}
 \includegraphics[width=3.5in]{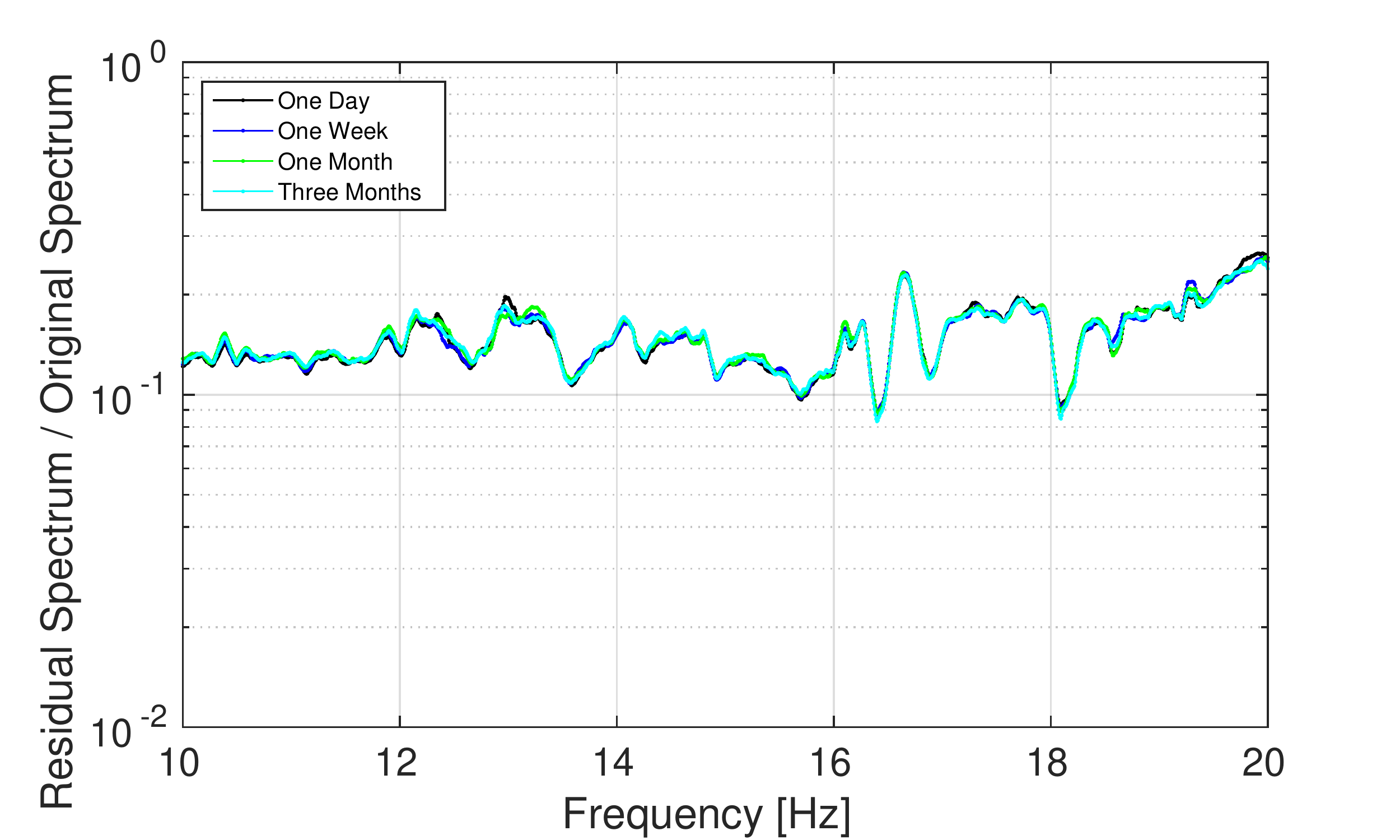} 
 \caption{Suppression of a seismometer (left) and tiltmeter (right) applying the same Wiener filter to different days after the day used for the calculation of the Wiener filter.}
 \label{fig:subtimeacc}
\end{figure*}

In Figure \ref{fig:subtimeacc}, we show the subtraction of seismometer and tiltmeter signals on a variety of timescales. Applying a filter calculated at some day to data recorded the same day, or days, weeks, months later, we see that subtraction performance is stable except for a signal around 19\,Hz in the seismometer case. A simple change in power of the 19\,Hz seismic source explains this result since subtraction of the 19\,Hz signal is limited by sensor noise. In general, variation in subtraction performance could also be due to changes of seismic correlations if dominant seismic sources change with time. In the tiltmeter case, no difference is seen.

\end{document}